\documentclass{aa}  

\usepackage{graphicx}
\usepackage{txfonts}
\usepackage{xcolor}
\usepackage{booktabs}
\usepackage[colorlinks=true,linkcolor=blue,allcolors=blue]{hyperref}
\usepackage{mdframed}

\newcommand{\lya}{Ly$\alpha$\xspace}
\newcommand{\logten}{\log_{10}}
\newcommand{\Llya}{L_{{\rm Ly}\alpha}}

\usepackage[normalem]{ulem}
\usepackage{color}
\definecolor{mygreen}{rgb}{0.0, 0.65, 0.20}

\defcitealias{Torralba-Torregrosa23}{TT23}

\definecolor{mycolor}{rgb}{0.122, 0.435, 0.698}
\definecolor{dscolor}{rgb}{0.000, 0.435, 0.698}

\definecolor{sgcolor}{rgb}{0.32941176470588235, 0.7058823529411765, 0.7764705882352941}

\newmdenv[innerlinewidth=1.0pt, roundcorner=6pt,linecolor=mycolor,innerleftmargin=6pt,innerrightmargin=6pt,innertopmargin=4pt,innerbottommargin=2pt]{mybox}

\begin{document} 

   % \title{The quasar \lya and UV luminosity functions at $2.7<z<5.3$ in the PAU Survey}
   \title{The PAU Survey: The quasar \lya and UV luminosity functions at $2.7<z<5.3$}
   \titlerunning{The PAU Survey -- \lya and UV luminosity functions of quasars}

   \author{
    Alberto Torralba-Torregrosa\inst{\ref{inst1}, \ref{inst2}}\thanks{E-mail: alberto.torralba@uv.es}
    \and Pablo Renard\inst{\ref{inst3}}
    \and Daniele Spinoso\inst{\ref{inst3}}
    \and Pablo Arnalte-Mur\inst{\ref{inst1}, \ref{inst2}}
    \and Siddhartha Gurung-López\inst{\ref{inst1}, \ref{inst2}}
    \and Alberto Fernández-Soto\inst{\ref{inst6}, \ref{inst7}}
    \and Enrique Gaztañaga\inst{\ref{inst8}, \ref{inst4}, \ref{inst5}}
    \and David Navarro-Gironés\inst{\ref{inst4}, \ref{inst5}}
    \and Zheng Cai\inst{\ref{inst3}}
    \and Jorge Carretero\inst{\ref{inst9}, \ref{inst10}}
    \and Francisco J. Castander\inst{\ref{inst4}, \ref{inst5}}
    \and Martin Eriksen\inst{\ref{inst11}, \ref{inst10}}
    \and Juan Garcia-Bellido\inst{\ref{inst12}}
    \and Hendrik Hildebrandt\inst{\ref{inst13}}
    \and Henk Hoekstra\inst{\ref{inst14}}
    \and Ramon Miquel\inst{\ref{inst11}, \ref{inst15}} 
    \and Eusebio Sanchez\inst{\ref{inst9}} 
    \and Pau Tallada-Crespí\inst{\ref{inst9}, \ref{inst10}} 
    \and Juan De Vicente\inst{\ref{inst9}}
    \and Enrique Fernandez\inst{\ref{inst11}}
    }

   \institute{
    Observatori Astron\`omic de la Universitat de Val\`encia, Ed. Instituts d’Investigaci\'o, Parc Cient\'ific. C/ Catedr\'atico Jos\'e Beltr\'an, n2, 46980 Paterna, Valencia, Spain\label{inst1}
    \and Departament d’Astronomia i Astrof\'isica, Universitat de Val\`encia, 46100 Burjassot, Spain\label{inst2}
    \and Department of Astronomy, Tsinghua University, Beijing 100084, China\label{inst3}
    \and Instituto de F\'{\i}sica de Cantabria (CSIC-UC), Avda. Los Castros s/n, 39005 Santander, Spain\label{inst6}
    \and Unidad Asociada ``Grupo de Astrof\'{\i}sica Extragal\'actica y Cosmolog\'{\i}a'', IFCA-CSIC/Universitat de Val\`encia, Val\`encia, Spain\label{inst7}
    \and Institute of Cosmology and Gravitation (University of Portsmouth), Portsmouth, UK\label{inst8}
    \and Institute of Space Sciences (ICE, CSIC), E-08193 Barcelona, Spain\label{inst4}
    \and Institut d'Estudis Espacials de Catalunya (IEEC), Edifici RDIT, Campus UPC, 08860 Castelldefels (Barcelona), Spain\label{inst5}
    \and Centro de Investigaciones Energ\'eticas, Medioambientales y Tecnol\'ogicas (CIEMAT), Avenida Complutense 40, E-28040 (Madrid), Spain\label{inst9}
    \and Port d'Informaci\'{o} Cient\'{i}fica (PIC), Campus UAB, C. Albareda s/n, 08193 Bellaterra (Barcelona), Spain\label{inst10}
    \and Institut de F\'{\i}sica d’Altes Energies (IFAE), The Barcelona Institute of Science and Technology, Campus UAB, 08193 Bellaterra (Barcelona), Spain\label{inst11}
    \and Instituto de Fisica Teorica (UAM/CSIC), Nicolas Cabrera 13, E-28049 Madrid, Spain\label{inst12}
    \and  Ruhr University Bochum, Faculty of Physics and Astronomy, Astronomical Institute (AIRUB), German Centre for Cosmological Lensing, 44780 Bochum, Germany\label{inst13}
    \and  Leiden Observatory, Leiden University, Einsteinweg 55, 2333 CC, Leiden, The Netherlands\label{inst14}
    \and Instituci\'o Catalana de Recerca i Estudis Avan\c{c}ats (ICREA), 0810 Barcelona, Spain\label{inst15}
    }

   \date{Accepted XXX. Received YYY; in original form ZZZ}

  \abstract{
  We present the Lyman-$\alpha$ (\lya) and ultraviolet (UV) luminosity function (LF), in bins of redshift, of quasars selected in the Physics of the Accelerating Universe Survey (PAUS). A sample of 915 objects was selected at $2.7<z<5.3$ within an effective area of $\sim 36$ deg$^2$ observed in 40 narrow-band filters (NB; FWHM $\sim 120$ \AA). We cover the intermediate-bright luminosity regime of the LF ($10^{43.5}<(\Llya/{\rm erg\,s}^{-1})<10^{45.5}$; $-29<M_{\rm UV}<-24$). The continuous wavelength coverage of the PAUS NB set allows a very efficient target identification and precise redshift measurements. We show that our method is able to retrieve a fairly complete ($C\sim 85\%$) and pure ($P\sim 90\%$) sample of \lya emitting quasars for $\Llya>10^{44}$ erg\,s$^{-1}$. In order to obtain corrections for the LF estimation, and assess the accuracy of our selection method, we produced mock catalogs of $0<z<4.3$ quasars and galaxies that mimic our target population and their main contaminants. Our results show a clear evolution of the \lya and UV LFs, with a declining tendency in the number density of quasars towards increasing redshifts. In addition, the faint-end power-law slope of the \lya LF becomes steeper with redshift, suggesting that the number density of \lya-bright quasars declines faster than that of fainter emitters. By integrating the \lya LF we find that the total \lya emitted by bright quasars per unit volume rapidly declines with increasing redshift, being sub-dominant to that of star-forming galaxies by several orders of magnitude by $z\sim 4$. Finally, we stack the NB pseudo-spectra of a visually selected ``golden sample'' of 591 quasars to obtain photometric composite SEDs in bins of redshift, enabling to measure the mean IGM absorption by the Lyman-$\alpha$ forest as a function of redshift, yielding results consistent with previous spectroscopic determinations.
  }


   \maketitle

%%%%%%%%%%%%%%%%%%%%%%%%%%%%%%%%%%%%%%%%%%%%%%%%%%%%%%%%%%%%%%%%%%%%
\section{Introduction}\label{sec:introduction}

Active galactic nuclei (AGN) are extremely energetic events powered by efficient accretion into super massive black holes (SMBH) in the cores of galaxies. These objects are among the most luminous in the Universe across the full electromagnetic spectrum \citep[e.g.,][]{Shen20}, and can be easily detected at very high redshift from ground-based observatories \citep[$z>6$; e.g.,][]{White03, Jiang16, Wang21}.
Recent studies confirmed that most massive galaxies host a SMBH in their center \citep[e.g.,][]{Askar22, Tremmel23}, and multiple lines of evidence have been found that these SMBH co-evolve with the bulge of the host galaxy \citep[e.g.,][]{Yang19, Mountrichas23}. Additionally, most galaxy formation models require feedback from AGNs to prevent the interstellar gas to overcool, quenching the star formation \citep[e.g.,][]{Dubois16, Dave19, Ward22} --- otherwise, the models predict an excessively large number of luminous galaxies, in contradiction with the observations \citep{Fabian12}. Hence, to investigate the properties of the AGN population, and how it evolves with cosmic time is a key piece to fully understand the big picture of galaxy evolution.

The luminosity function (LF) of AGNs, defined as the number density of these objects per comoving volume as a function of luminosity, evolves with cosmic time \citep[e.g.,][]{Kulkarni19, Shen20}, both in shape and normalization: the-faint end of the bolometric AGN LF peaks at $z\approx 2$--$2.5$ -- coinciding with the peak of star formation history \citep[e.g.,][]{Madau14} whilst the bright-end peaks at slightly earlier cosmic time $z\approx 3$, and evolves slower toward the lowest redshifts, in what is often called ``cosmic downsizing'' \citep[e.g.,][]{Hasinger05, Hopkins07, Croom09, Fanidakis12}.

The Lyman-$\alpha$ (\lya ) emission-line is typically the most luminous in the rest-frame ultraviolet (UV) spectra of quasars \citep[QSO; e.g.,][]{Vanden-Berk01, Ivaschenko14, Harris16}, and star forming galaxies \citep[SFG; e.g.,][]{Partridge67, Pritchet94, Nakajima18}. The observed wavelength of \lya is redshifted into the optical range at $z\sim 2$--6, allowing to detect this line in ground-based  surveys. Sometimes, secure identification of \lya is possible even without an explicit detection of the spectral continuum \citep[e.g.,][]{Bacon15}.
On the one hand, the faint-end of the \lya LF is dominated by a population of typically small, young, metal and dust-poor SFGs \citep[$\Llya\lesssim 10^{43.5}$ erg\,s$^{-1}$; e.g.,][]{Guaita10, Drake17, ArrabalHaro20, Santos20}. The \lya LF deviates from a typical Schechter shape for $\Llya\gtrsim 10^{43.5}$ erg\,s$^{-1}$, and the detection of X-ray counterparts of the sources at this range of \lya luminosities suggests that most of them are powered, at least partially, by AGNs \citep{Sobral18, Matthee17}.
On the other hand, the brightest end of the \lya LF ($\Llya > 10^{43.5}$ erg\,s$^{-1}$) is dominated by QSOs \citep[e.g.,][]{Calhau20, Spinoso20}.
The \lya LF in the whole range is well described by a double Schechter function \citep{Schechter76}.
The intermediate \lya luminosity regime ($\Llya\sim 10^{43.5}$ erg\,s$^{-1}$) is particularly interesting, because it might contain objects where the \lya emission from the AGN and star-formation activity is similarly relevant \citep[e.g.,][]{Sobral18}.
While the \lya LF of SFGs has been widely explored in the literature, \citep[e.g.,][]{Cowie98, Hu98, Martin04, Malhotra04, Dawson07, Ouchi08, Blanc11, Adams11, Bacon15, Karman15, Drake17, Santos16, Konno18, deLaVieuville19, Hu19, Khusanova20, Ono21, Santos21, Morales21, Wold22, Xu23, Thai23}, the AGN-dominated, bright-end of the \lya LF has only been constrained in the recent years \citep{Spinoso20, Zhang21, Liu22b, Torralba-Torregrosa23}.

In the same manner as the \lya LF, the UV LF is dominated by SFG at the faint-end  ($M_{\rm UV} \gtrsim -23$) and AGN in the bright-end \citep[$M_{\rm UV} \lesssim -23$; e.g.,][]{Adams23}. Contrary to the \lya LF, the bright-end of the UV AGN LF has been broadly explored \citep[e.g.,][]{Glikman11, Ross13, McGreer13, Palanque-Delabrouille16, Jiang16, Yang16, Kulkarni19, Akiyama18, Matsuoka18, Schindler19, Niida20, Pan22}. However, the faintest end of the AGN UV LF is still uncertain. In particular, a population of faint dust-obscured AGNs has been unveiled by the most recent JWST observations \citep[e.g.,][]{Labbe23, Matthee24, Kokorev24, Kocevski24, Greene24, Akins24}, whose number density appears to be larger by more than one order of magnitude than the expected when extrapolating the bright-end of the AGN UV LF. The connection between this population of faint dusty AGNs and bright QSOs is still unclear, and the determination of the AGN LF in the intermediate luminosity regime ($M_{\rm UV}$ between $-24$ and $-20$) is crucial in order to solve this problem. The global determination of the AGN LF is key to determine the role of this population in the cosmic reionization. Hydrodynamical and radiative transfer simulations suggest that the contribution of AGNs to hydrogen reionization is sub-dominant \citep[e.g.,][]{Dayal20, Trebitsch21}. This is supported by the rapid decline of the number of bright quasars at high redshifts \citep[e.g.,][]{Kulkarni19}. However, the role of AGNs can be determinant in helium reionization, which happened at later cosmic time \citep[e.g.,][]{Compostella14, Worseck16}.

In this context, in this work we seek to identify and characterize the population of quasars with bright ($\Llya>10^{43.5}$ erg\,s$^{-1}$) \lya emission using the narrow-band (NB) data from the Physics of the Accelerating Universe Survey (PAUS). Multi-NB surveys such as PAUS are extremely suitable for this task. On the one hand, a continuous coverage of great part of the optical range with NB photometry yields a low-resolution spectrum for every source in the field-of-view without any target pre-selection, allowing for very complete samples, if carefully selected. On the other hand, it becomes possible to probe a wide and continuous range of redshift, unlike surveys that only use a few NBs placed in strategic wavelengths. In addition, the availability of pseudo-spectra can be used to obtain precise photometric redshifts \citep{Marti14, Eriksen20, Alarcon21, Hernan-Caballero21, Laur22, Navarro-Girones23, Hernan-Caballero24}.

In \cite{Torralba-Torregrosa23} \citepalias[hereafter][]{Torralba-Torregrosa23}, we presented a method to select \lya emitters in the miniJPAS \citep{Bonoli21} and J-NEP \citep{Hernan-Caballero23} multi-NB data. In this work we adapt and improve the selection method to the PAUS data, and use it to estimate the UV and \lya LFs of QSOs selected through \lya emission. We also use the obtained sample of \lya-selected QSOs to estimate the mean absorption by the intergalactic medium (IGM) due to the \lya forest as a function of redshift \citep[e.g.,][]{Faucher-Giguere08, Becker13, Inoue14}.

This paper is structured as follows. In Sect.~\ref{sec:observations} we describe the observational data of the PAU Survey, employed throughout this work. In Sect.~\ref{sec:Methods}, we first describe the mock catalogs, used to test and validate our selection method; we define our \lya selection method, and the procedures to estimate the photometric redshifts of the sources, the \lya luminosity and to obtain the \lya and UV LFs. In Sect.~\ref{sec:results_and_discussion} we present the \lya-selected sample, and produce photometric composite QSO spectra in bins of redshift. We present the \lya and UV LFs, and their integrated values as a function of redshift. In Sect.~\ref{sec:discussion} we compare our measured \lya and UV LFs with previous determinations in the literature, and discuss the evolution of these with redshift. We also comment on the efficiency of our selection method, comparing with other AGN samples in the literature, and cross-matching our catalog with spectroscopic surveys. Finally, Sect.~\ref{sec:summary} summarizes the contents of this work.

Throughout this work we use a $\Lambda$CDM cosmology as described by \cite{Planck18}, with $\Omega_\Lambda=0.69$, $\Omega_\text{M}=0.31$, $H_0=67.7$ km\,s$^{-1}$\,Mpc$^{-1}$; unless specified otherwise. All the magnitudes are given in the AB system \citep{Oke83}.

%%%%%%%%%%%%%%%%%%%%%%%%%%%%%%%%%%%%%%%%%%%%%%%%%%%%%%%%%%%%%%%%%%%%
\section{Observations: The PAU Survey}\label{sec:observations}

The \textit{Physics of the Accelerating Universe Survey} (\textit{PAUS}) is a wide-field multi-narrowband survey carried out by the PAUCam instrument \citep{Padilla19}, mounted at the 4.2 m \textit{William Herschel} Telescope (WHT) at the Roque de los Muchachos Observatory in the Canary Islands. PAUCam is a 18-detector camera, with a FoV of diameter $\sim 1^\circ$, designed to cover a wide area of the sky making use of a filter set composed of 40 NBs (FWHM $\sim 115$ \AA). The unique set of NB filters provides continuous wavelength coverage over observed wavelengths 4500--8500 \AA , allowing to obtain a low resolution pseudo-spectrum for every source in the FoV. The detection of diffuse \lya emission in PAUS has already been evaluated in \citet{Renard2021}, and other prominent spectral features have already been studied with PAUS data, such as emission line ratios \citep{Alarcon21}, or the D4000 spectral break \citep{Renard2022}.

PAUS has targeted the CFHTLS \citep{Cuillandre12} W1, W3 and W4 fields; the GAMA G09 field \citep{Driver22}; and the COSMOS field \citep{Scoville07}. In this work we use the data of W1, W3 and G09 (hereafter dubbed W2). In total, these three fields are covered with the full set of 40 filters in an effective area of 35.68 deg$^{2}$ (after masking). For PAUS catalogs, all objects with magnitudes brighter than $i=23$ are selected; at fainter magnitudes, the photometry S/N is too low to reliably measure galaxy redshifts \citep{Alarcon21, Navarro-Girones23}.

The standard PAUS photometry is computed using forced photometry \citep{Serrano2022}. For each object and narrow-band, its fluxes are separately measured on each single-exposure image. Three exposures are taken per pointing, though the number of exposures on individual objects may vary if they are positioned near the edges of the tiling, due to overlaps and dithering.
The flux apertures are ellipses that contain 62.5\% of the flux, according to their Sérsic profiles convolved with the seeing of each exposure; the total flux for a given object and narrow band is the inverse-variance weighted average of all exposures. This method results in a custom aperture for each object, fitted to their size and morphology, which in turn yields a higher S/N than standard fixed apertures \citep{Serrano2022}.
For this work, we computed custom photometric runs of W1, W2 and W3 utilizing the same data reduction outlined in \citet{Serrano2022}, but using fixed circular apertures of diameter 3\arcsec \, instead. This fixed aperture was chosen in order to be consistent with the photometry used in \citetalias{Torralba-Torregrosa23}.

The positions of all objects, their $i$ magnitudes and morphological parameters are extracted from broad-band reference catalogs; no detections are performed on PAUS narrow-band images. The reference catalog used for the W1 and W3 fields is the CFHTLS data release from \citet{Heymans2012}, and for the W2 (GAMA G09) field it is the fourth KiDS data release \citep{Kuijken19}.

%%%%%%%%%%%%%%%%%%%%%%%%%%%%%%%%%%%%%%%%%%%%%%%%%%%%%%%%%%%%%%%%%%%%
\section{Methods}\label{sec:Methods}

In this Section we describe the methodology employed to select high redshift QSOs in the PAUS wide fields, and the estimation of the \lya and UV luminosity functions. We first describe the mock catalogs developed to mimic our observations, which we use to test and assess the effectiveness of our methods (Sect.~\ref{sec:mocks}). Next, we describe our target selection method, based on \lya emission detection (Sect.~\ref{sec:selection_method}), we characterize the purity and completeness of our method (Sect.~\ref{sec:puricomp}) and describe the procedure to estimate the \lya and UV luminosity functions (Sect.~\ref{sec:lya_LF_computation}).

\begin{figure*}
    \centering
    \includegraphics[width=\linewidth]{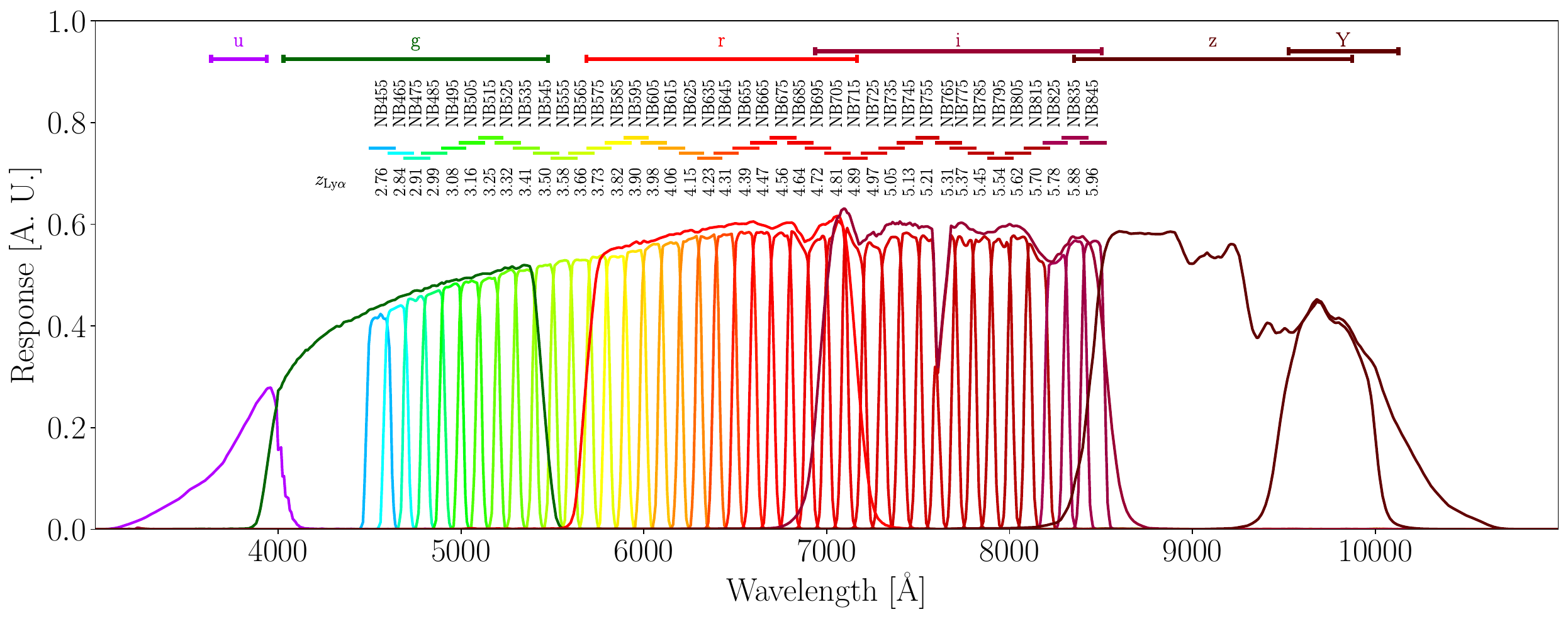}
    \caption{Transmission curves of PAUS NBs and BBs of the reference catalogs. On top of the transmission curves, horizontal lines are displayed showing the FWHM of each filter. We show the \lya redshift corresponding to the pivot wavelength of each NB. As shown in Table~\ref{tab:NB_lya_redshifts}, only NB455-NB755 are used in this work, since we obtained no reliable detections in NBs of higher wavelengths.}
    \label{fig:filter_transmission_curves}
\end{figure*}

\subsection{Mock catalogs}\label{sec:mocks}

We generate mock catalogs and photometric NB pseudo-spectra of \lya emitting QSOs (our target sample) and their common contaminants (low-$z$ QSOs and galaxies), with the aim of characterizing the effectiveness of our selection method. The mocks are employed to train machine learning algorithms to improve the selection function (Sect.~\ref{sec:selection_method}) and to measure accurate photometric redshifts (Sect.~\ref{sec:redshift_correction}); to estimate the purity and completeness of the selected sample (Sect.~\ref{sec:puricomp}) and to obtain corrections for the estimation of the \lya and UV luminosity functions (Sect.~\ref{sec:lya_LF_computation}).

\begin{figure}
    \centering
    \includegraphics[width=\linewidth]{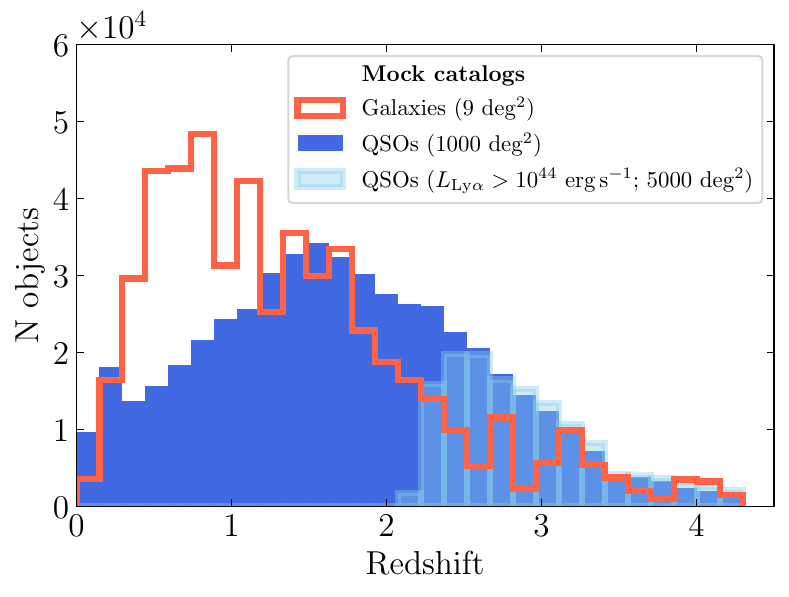}
    \caption{Redshift distribution of galaxies within the QSO and low-$z$ galaxy mocks. The different mock populations were simulated in different areas: 9, 1000 and 5000 deg$^2$ for the low-$z$ galaxy, QSO and high $\Llya$ QSO mocks, respectively. In all mocks we applied conservative magnitude cuts of $r<25$. Despite the much larger area used in the QSO mocks, the low-$z$ galaxies are significantly more numerous.}
    \label{fig:lowz_mock_redshift_distribution}
\end{figure}

\subsubsection{QSO mock catalog}\label{sec:qso_mock}

The mock catalog of the QSO population at $0<z<4.5$ is generated following the same procedure as in \citetalias{Torralba-Torregrosa23}, which at the same time is based on the method presented in \cite{Queiroz22} for a QSO mock population in the J-PAS survey. We select spectra in this redshift range from the SDSS DR16 SupersetQ \citep{Lyke20} with good median signal-to-noise over all pixels (\texttt{SN\_MEDIAN\_ALL}>0), measured redshift with high level of confidence (\texttt{ZWARNING}=0); and classified as QSO (\texttt{IS\_QSO\_FINAL}=1). Next, we sample values over a grid of redshift and $r$-band magnitude from the luminosity functions of \cite{Palanque-Delabrouille16} (PLE+LEDE model). We look for the closest objects in redshift within the SDSS DR16 SupersetQ sub-sample, and apply the corresponding multiplicative factors to the observed wavelength array and fluxes to match the sampled values of ($z,r$). In \citetalias{Torralba-Torregrosa23} we show that this procedure yields a population of QSOs that, for $z>2$ is in good agreement with the \lya LF measured in the literature (see Fig.~2 in \citetalias{Torralba-Torregrosa23}). We produce a QSO mock with a size equivalent to 1000 deg$^2$, and a $\times5$ larger area for objects with $\Llya>10^{44}$ erg\,s$^{-1}$ to overcome the limitations due to the low number of objects at such high luminosities. The sizes of the simulated areas are chosen as arbitrarily large values, enough to obtain smooth distributions in the mock properties (for more details about the QSO mock see \citetalias{Torralba-Torregrosa23}). The number of objects in these mocks as a function of redshift is shown in Fig.~\ref{fig:lowz_mock_redshift_distribution}.

As a final step, we compute synthetic photometry of all the selected objects using the transmission curves of all the PAUS filters to obtain pseudo-spectra. Since all the SDSS spectra used to generate the mocks are selected to have good S/N, the error of our synthetic photometry is negligible compared to the error of real PAUS observations. We model the PAUS photometric errors separately. For this, we model the  uncertainties as a function of photometric magnitude with an exponential (see \citetalias{Torralba-Torregrosa23}), fitting the exponential parameters for every PAUS filter in each one of the W1, W2 and W3 fields. We add random Gaussian perturbations to the synthetic fluxes of the mock according to this model to obtain the final mock QSO catalog.

\subsubsection{Low-redshift galaxy mock catalog}

Among the main contaminants in the search for \lya emission there are nebular emission-lines associated to star-formation such as H$\alpha$ ($\lambda\,6565$ \AA), H$\beta$ ($\lambda\,4861$ \AA), [\ion{O}{III}] ($\lambda\lambda\,4959,5007$ \AA) or [\ion{O}{II}] ($\lambda\lambda\,3727,3729$ \AA) emitted by low-$z$ galaxies (i.e., $z<2$, in this context). On top of their eventual bright line-emission, low-$z$ galaxies may contaminate our LAE sample due to random fluctuations in their NB fluxes. This source of contamination may be relevant because low-$z$ galaxies account for the bulk of the objects in the parent photometric catalog. Despite this, \citetalias{Torralba-Torregrosa23} found little contamination of nebular emission lines form $z<2$ galaxies, with the minor exception of a few extreme [\ion{O}{II}] emitters.

In order to assess the impact of low-$z$ galaxies as contaminants in our samples, we make use of a galaxy mock, built by using the semi-analytic model (SAM) \texttt{L-Galaxies} on top of the dark-matter merger trees of the \texttt{Millennium} N-body dark matter (DM)-only simulation \citep{Springel05}. In particular, the version of \texttt{L-Galaxies} we use is derived from the public release of \cite{henriques2015}, with the inclusion of detailed models for the formation, mass-growth and dynamical evolution of SMBHs recently presented in \cite{izquierdo-villalba2020,izquierdo-villalba2022a} and \cite{spinoso2023}.

On top of being one of the most detailed SAMs in the literature to describe the evolution of galaxies and SMBHs, \texttt{L-Galaxies} includes the possibility to produce photometric mock-data on-the-fly. Indeed, \cite{izquierdo-villalba2019mocks} presented a methodology to simulate an observed light-cone for a generic multi-band photometric survey, showing the example of the J-PLUS survey \citep{Cenarro19}. In particular, their method relies on simulating the continuum emission of galactic stellar populations by means of a galaxy-templates grid which depends on metallicity, stellar age and a given choice for the stellar initial mass function \citep[\citealt{izquierdo-villalba2019mocks} opted for a standard Chabrier IMF;][]{Chabrier2003}. On top of the stellar continua, nine different nebular emission lines\footnote{Namely:
Ly$\alpha$ (1215.67 \AA), H$\beta$ (4861 \AA), H$\alpha$ (6563 \AA), [$\rm O\,II$] (3727 \AA, 3729 \AA), [$\rm O\,III$] (4959 \AA, 5007 \AA), [$\rm Ne\,III$] (3870 \AA), [$\rm O\,I$] (6300 \AA), [$\rm N\,II$] (6548 \AA, 6583 \AA), and [$\rm S\,II$] (6717 \AA, 6731 \AA).} from star-forming regions are simulated by using the method presented in \cite{orsi2014}, where the emission line flux is obtained as a function of stellar age, gas density, metallicity and ionization parameter \citep[see][for further details on the implementation of the emission-lines model in \texttt{L-Galaxies}]{izquierdo-villalba2019mocks}.

In order to build our low-$z$ galaxy mock we follow the methodology outlined above, adapting it to the PAUS photometric filter-set and survey volume. In particular, we simulate a $\rm 3\times3\,deg=9\,deg^2$ sky patch; that is, roughly one-fourth of the effective PAUS area we use (see Sect. \ref{sec:parent_sample}). The choice of this area is mainly limited by computation time for the lightcone. Despite the much smaller area as compared to the QSO mock, the number of galaxies in the lightcone is similar to the number of objects in the QSO mock (see Fig.~\ref{fig:lowz_mock_redshift_distribution}). We apply a conservative magnitude cut (i.e., $i<25$) to the photometry of galaxies within the light-cone. Given the effective depth of PAUS data, this magnitude cut allows us to significantly reduce the size of our low-$z$ galaxy mock data without compromising their completeness. The PAUS photometric errors are simulated as described in Sect.~\ref{sec:qso_mock} for the QSO mock. Our final mock catalog contains $\sim 5\times10^{5}$ galaxies with $i<25$ at redshifts $0<z\lesssim6$ (median $z=1.2$) (see Fig.~\ref{fig:lowz_mock_redshift_distribution}).

\subsection{Lyman-$\alpha$ selection method}\label{sec:selection_method}

In this Section we describe the selection method used to obtain a sample \lya selected sample of QSOs in the footprint of PAUS.

\subsubsection{Parent sample}\label{sec:parent_sample}

We use the diameter 3\arcsec\ aperture photometry catalogs of the PAUS wide fields W1, W2 and W3 as discussed in Sect.~\ref{sec:observations}. For our parent sample, we use only the regions of these fields where there is coverage for all 40 PAUS NBs. The reference catalogs used by these fields are taken from surveys with deeper broad-band imaging and better PSF than the average of PAUS NBs. Due to this, some sources from the reference catalogs are blended in the PAUS NB images, leading to contamination in the fixed aperture photometry. Hence, we remove objects with neighbours with a separation closer than 3\arcsec. This cut removes the 8\% of the objects across the W1, W2 and W3 catalogs, hence this will be taken into consideration in the computation of the LF.

After removing potentially blended sources, and applying the PAUS field masks for the area covered by all 40 NB filters, we are left with a parent sample containing 1\,022\,196 objects (239\,312, 267\,393 and 515\,491 in W1, W2 and W3, respectively) in a total area of 35.7 deg$^2$ (effective area after masking).

\begin{figure}
    \centering
    \includegraphics[width=\linewidth]{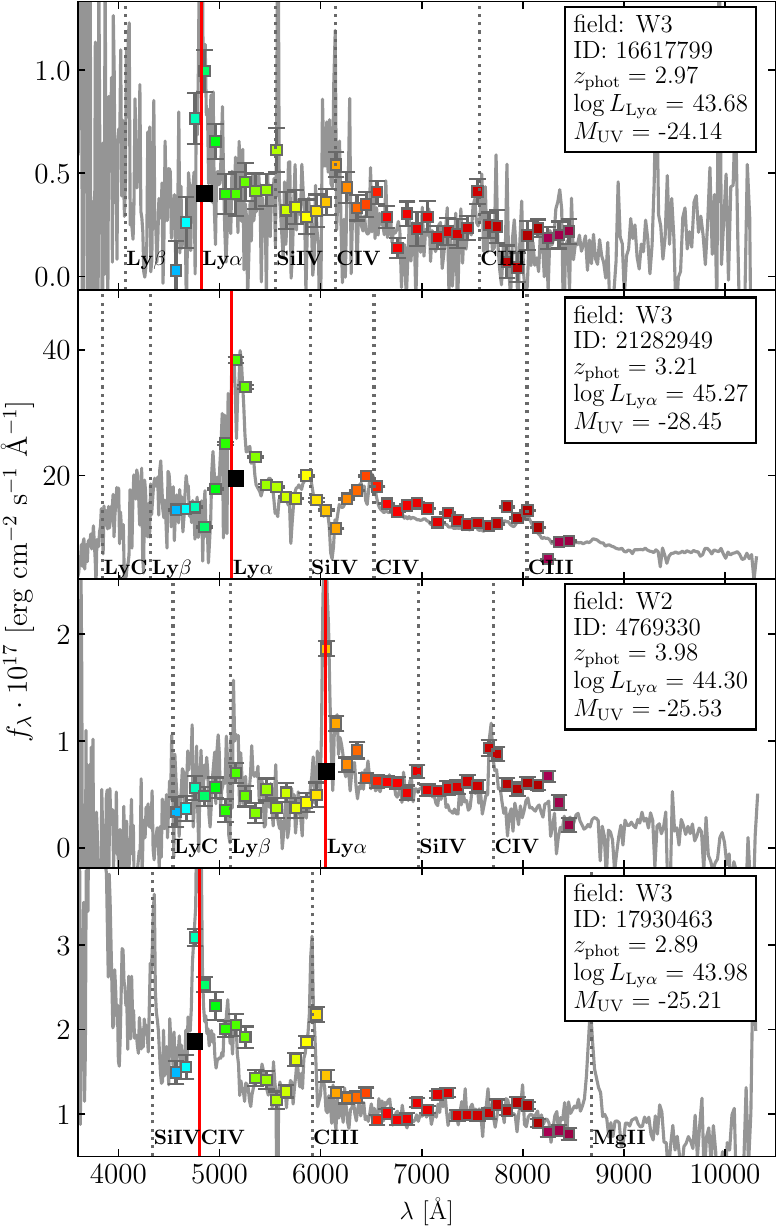}
    \caption{Four examples of candidates selected by our pipeline. The colored squares represent the NB fluxes of PAUS. We compare with the SDSS spectrum in gray (resampled by a factor 10 to reduce visual noise). We mark the most important QSO emission lines (dotted gray vertical lines) and the predicted observed wavelength of \lya by our selection pipeline (vertical red line). The estimation of the continuum level is marked with a black square. The first three panels are examples of correctly selected \lya emitters. The last panel is an example of a confusion between \lya and \ion{C}{IV} ($z_\ion{C}{IV}=2.1$), the most common source of contamination.}
    \label{fig:example_candidates}
\end{figure}

\subsubsection{NB flux excess selection}\label{sec:NB_excess}

We search for \lya emission in each of the PAUS narrow bands through a NB flux excess selection. Each narrow band probes \lya lines at redshifts within $\delta z = \pm 0.06$ around the central redshifts listed in Table~\ref{tab:NB_lya_redshifts}, and displayed in Fig.~\ref{fig:filter_transmission_curves}. This redshift interval is defined by the FWHM of the NB filters. A detailed description of the LAE candidate selection method can be found in \citetalias{Torralba-Torregrosa23}.
In the following Sections we summarize the method, which was adapted for the particularities of PAUS data.

In the first place, we look for tentative \lya detections in all the NBs. For each NB, an estimation of the continuum flux is obtained by averaging the photometric fluxes of the 5 closest NBs to the one of interest, at bluer and redder wavelengths, ignoring both directly adjacent NBs to avoid contamination from the tail of broad \lya lines \citep[see e.g.,][]{Greig17}. That makes a total of 10 NB filters to compute the continuum level. In addition, the flux of the NBs blueward from \lya is affected by the \lya forest. The mean absorption of the \lya forest is corrected in the continuum estimation using the curve for the mean IGM transmission in \cite{Faucher-Giguere08}. After the estimation of the continuum flux level of every source in our parent sample, we look for NBs with S/N $>6$ and a 3$\sigma$ flux excess over the continuum:
\begin{equation}
f_{\rm NB}^\lambda - f_{\rm cont}^\lambda > 3\sqrt{\sigma_{\rm NB}^2 + \sigma_{\rm cont}^2}\,.
\end{equation}
We also impose that this excess is compatible with a \lya line with ${\rm EW}_0 \geq {\rm EW}_0^{\rm min} = 20$\ \AA{} by applying
\begin{equation}
\frac{f^\lambda_{\rm NB}}{f^\lambda_{\rm cont}}
> 1 + \frac{(1 + z_{\rm NB}) \cdot {\rm EW}_0^{\rm min}}{\rm FWHM_{NB}}\,,
\end{equation}
where $z_{\rm NB}$ is the \lya redshift corresponding to the target NB. (see \citetalias{Torralba-Torregrosa23}, also e.g., \citealt{Santos16, Sobral18, Spinoso20}). If two or more contiguous NBs meet these criteria, the \lya detection is assigned to that of higher flux.

After the preliminary \lya detection, we similarly search for other lines in the selected sources. For this secondary line search, following the optimal search criteria in \citetalias{Torralba-Torregrosa23}, we look for NBs with $>3\sigma$ flux over continuum excess and EW$_{\rm obs} > 100$ \AA . These emission lines are searched using a continuum estimate without the \lya forest transmission correction. The algorithm checks the secondary line detections and discards candidates with lines not compatible with the expected position of the most prominent QSO lines other than \lya ; in particular Ly$\beta$+\ion{O}{VI}, \ion{C}{IV}, \ion{C}{III}] and \ion{Mg}{II}.
Figure \ref{fig:example_candidates} shows four hand-picked examples of selected candidates using the procedure described above. The first three panels show correctly selected QSOs with coinciding spectroscopic \lya redshifts of their SDSS counterparts. In these three examples, we also detect the secondary \ion{C}{IV} line, which ensures a correct determination of the redshift. The bottom panel shows an example of the most common contaminant: a confusion between \lya and a lower redshift ($z=2.3$) \ion{C}{IV} emission line.

\subsubsection{Artificial neural network classifier}\label{sec:ML_class}

In \citetalias{Torralba-Torregrosa23} a visual inspection was performed through all the candidates in order to remove obvious contaminants such as low-$z$ galaxies with extended morphology, or NBs affected by cosmic rays and other artifacts. In this work, we trained a multi-layer perceptron artificial neural network (ANN) model in order to identify secure candidates and discard contaminants. This is motivated by the notable increase in the size of the candidate sample with respect to that in \citetalias{Torralba-Torregrosa23}, and the potential application of the method to even larger datasets in the future.

In order to train the ANN, we generate training and test sets from the subsets of objects in our mocks selected by NB flux excess (see Sect.~\ref{sec:NB_excess}). Each set contains equal number of the three classes to consider: (1) QSO with the correct narrow-band redshift, (2) QSO with wrong narrow-band redshift and (3) low-$z$ galaxy. The first of these three classes corresponds to the true positive LAE detections, while the other two represent the main contaminants. 

\begin{figure}
    \centering
    \includegraphics[width=\linewidth]{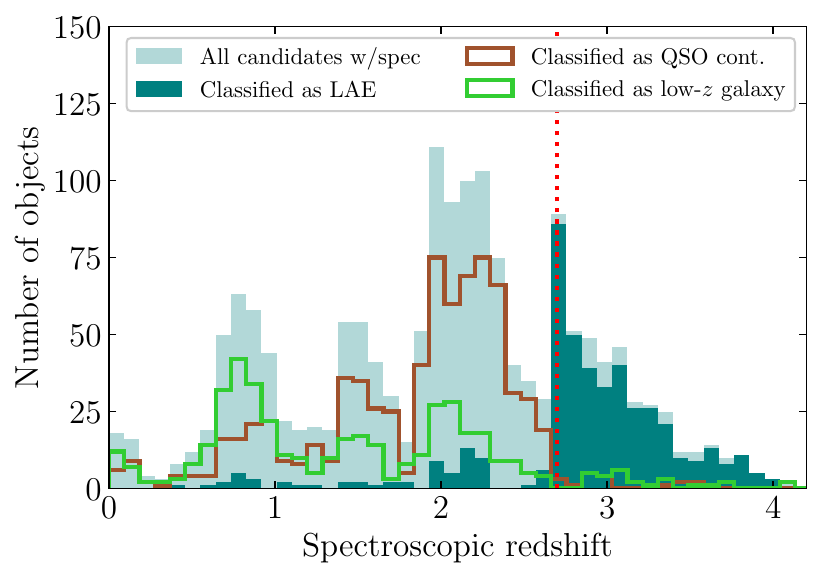}
    \caption{Classification of the candidates with spectroscopic counterpart by our ANN algorithm. The solid light teal histogram represents the full spectroscopic sample, and the solid dark blue the sub-sample classified as LAE by the NN. The empty brown and green histograms represent the candidates classified as contaminant QSOs and low-$z$ galaxies, respectively. For our target redshift range ($z\gtrsim2.75$), the algorithm correctly classifies most objects ($\sim 90\%$ completeness). The red dotted vertical line marks the minimum redshift targeted by the NB with the lowest wavelength ($z=2.7$, NB455).}
    \label{fig:ML_spec_hist}
\end{figure}

The input for the ANN classifier is composed of 46 features:
\begin{itemize}
    \item All 40 NB fluxes. Normalized to the maximum NB flux of each object.
    \item $g,r,i$ and $z$ BB fluxes. Normalized to the maximum BB flux. The band $u$ is omitted for being partially out of the SDSS range, used to create the mocks.
    \item $r_{\rm synth}$. Synthetic broad-band magnitude computed from the NB whose wavelength coverage reproduces the classical $r$ band ($\lambda\approx 5750$--7000 \AA). Normalized with a factor $1/30$ (so every value is $<1$).
    \item NB$_\text{\lya}$: the NB of the \lya detection, numbered from 0 to 39. Normalized with a factor $1/39$.
\end{itemize}

The NB and BB arrays of fluxes are normalized separately since the BB data is obtained from reference catalogs external to PAUS (see Sect.~\ref{sec:observations}), and their fluxes might not be directly comparable to PAUS 3\arcsec aperture NB fluxes. The $r_{\rm synth}$ feature gives information about the apparent magnitude of the object, and NB$_\text{\lya}$ establishes an initial guess of the redshift of the source. A synthetic composite $r_{\rm synth}$ broad-band is used instead of $r$ from the reference catalogs for consistency.

We perform a search for optimal parameters of the ANN over a wide parameter space. The optimal parameters obtained in the grid-search, and more details about the ANN architecture can be found in Appendix~\ref{sec:NN_class_details}. As shown in Fig.~\ref{fig:conf_matrix_NN}, the classifier yields a $\sim 90\%$ pure sample of QSO LAEs over the test set ($\sim 95\%$ for $\Llya>10^{44}$ erg\,s$^{-1}$), with 7\% (2\%) contamination from QSOs (galaxies) at lower redshifts. The completeness of the sample classified as QSO LAE is 96\% and 98\% in the test set, for the general sample and for $\Llya>10^{44}$ erg\,s$^{-1}$, respectively.

We also determine the accuracy of the ANN classifier with our spectroscopic sub-sample (presented in Sect.~\ref{sec:Xmatch_spec} below). In Fig.~\ref{fig:ML_spec_hist} we show the classification over this sub-sample. For $z\gtrsim 2.75$, our target redshift range (see Table~\ref{tab:NB_lya_redshifts}), the classifier correctly identifies as LAEs $\sim 90\%$ of the objects, in line with the completeness estimated in the test set. For lower redshift, $z\lesssim 2.75$ (i.e., the range of the contaminants), the algorithm successfully identifies most of the objects as contaminants. The confusion between contaminant QSO and galaxies is irrelevant in this work, since we are only interested in the QSO LAE sub-sample. The purity derived from this result is $\sim 86\%$ ($>90\%$ for $\Llya>10^{44}$ erg\,s$^{-1}$), again in line with that of the test set.

\subsubsection{Visual inspection}\label{sec:visual_inspection}

The mock data used to train the ML model described in Sect.~\ref{sec:ML_class} only contains objects with clean synthetic photometry, performed over either simulated SEDs (in the case of the low-$z$ galaxy mock) or spectra with no warnings raised by the SDSS pipelines (in the case of the QSO mocks). Hence, the training set does not account for objects affected by instrumental artifacts, cosmic rays or photometry calibration problems, among other issues that can affect the NB fluxes leading to a false detection. For this reason, despite the good performance shown by the ML model, a visual inspection is still necessary in order to catch objects misclassified by the algorithm due to unexpected features. Nonetheless, the ML classification significantly reduced the candidate sample for the visual analysis, from an initial sample of 4356 objects to 999 candidates.

The NB photometry of some objects is clearly affected by the extended emission of a nearby object, leading to unusually high NB fluxes that are mistaken by \lya emission lines. In some other objects, the images were affected by cosmic rays, instrumental artifacts or similar effects that alter the NB fluxes. Finally, some images of the PAUS fields are affected by calibration issues, resulting in unusually large zero-points. The zero-points are computed with a cross-match of bright SDSS stars on a per-CCD basis \citep[][each CCD zero-point being the median of the reference stars]{Serrano2022}. Hence, flux underestimation of reference stars (e.g., due to scatter-light or cross-talk between CCDs) could result on abnormally high zero-points for a few exposures. This issue artificially enhances the measured flux of some NBs in some regions, leading to wrong \lya line detections. All these problems cannot be trivially reproduced in our mocks, thus the affected objects must be removed from the selected sample via visual classification.

We visually inspected the initial sample of 999 QSO LAE candidates selected by our method, and agreed on the final classification. The inspection yielded a ``golden sample'' of 591 (59\%) objects selected as true positive QSO LAEs at the correct redshift. On the other hand, a non-negligible amount of 102 (10\%) objects were removed from the final sample due to issues with their pseudo-spectra or images. Some of these removed objects showed contamination by cosmic rays in their selected NB, evident instrumental artifacts, extended emission from nearby extended objects or clear issues with the photometric calibration that led to unusual pseudo-spectra. The remaining 316 (31\%) objects are are left as ``uncertain'', not being classified as true positives nor as contaminants.

\subsubsection{High-$z$ sample}\label{sec:high_z_sample}

In addition to the search of candidates in the interval $2.7<z<4.2$, we look for objects at higher redshifts. However, since our machine learning algorithm was trained using SDSS data, it is limited to the above mentioned redshift range, given that the SDSS sample becomes too scarce for higher redshifts. We perform a candidate search for $z>4.2$, without applying the ANN classification, and inspect all selected objects to identify secure \lya emitting QSOs. This search produced a sample of 22 visually selected QSOs ($\sim 0.6$ deg$^{-1}$) at $4.2\lesssim z \lesssim 5.3$ (see Table~\ref{tab:NB_lya_redshifts}), from an initial sample of 549 high-$z$ candidates.

\subsubsection{NB redshift correction}\label{sec:redshift_correction}

\begin{figure}
    \centering
    \includegraphics[width=\linewidth]{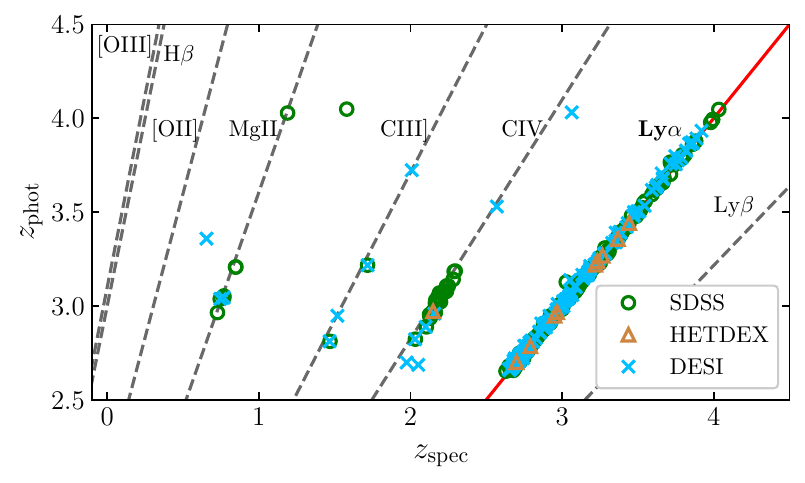}
    \includegraphics[width=\linewidth]{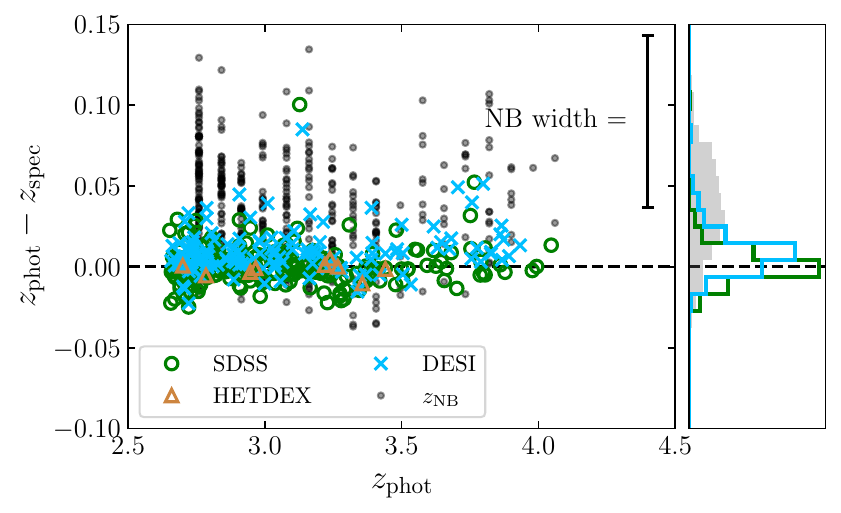}
    \caption{{\bf Top:}
    Comparison between photometric and spectroscopic redshifts of the candidates in our sample with counterparts in SDSS DR18 (green circles) HETDEX (orange triangles) and DESI DR1 (blue crosses). The straight lines mark the confusion between \lya and other bright QSO lines. Thus, objects laying on the red line are true positive detections of LAEs. {\bf Bottom:} Difference between the photometric and spectroscopic redshifts of the objects shown in the top panel. The black dots represent the same objects using the photometric redshift of the selection NB, before applying the machine-learning algorithm presented in Sect.~\ref{sec:redshift_correction}. The colored dots represent the same objects after the random forest redshift correction, for each spectroscopic subsample. The bottom right panel shows normalized histograms, result of collapsing the left panel in the horizontal axis. The HETDEX histogram is not shown due to poor statistics.}
    \label{fig:redshift_phot_spec}
\end{figure}

Each NB can probe an observed \lya wavelength in an interval defined by the width of the transmission curve. The mean redshift probed by the PAUS NBs used in this work are summarized in Table~\ref{tab:NB_lya_redshifts}. In addition, the full 40-NBs pseudo-spectra provide information that can be used to better constrain the redshifts of the detected sources. This is especially relevant for the correct assessment of the redshift of sources selected via \lya detection. The \lya emission-lines of these sources typically present broad profiles \citep[width $\sim 4500$ km/s;][]{Greig16}, and the absorption due to the \lya forest (for $\lambda_{\rm obs}<\lambda_{\rm obs}^\text{\lya}$) typically induces a bias in the \lya redshift towards slightly higher values.

We improve the redshift accuracy of our selected objects using machine learning algorithms. In order to obtain train/test sets, we run our selection method on our mocks. Then we keep only those candidates whose difference between the NB redshift ($z_{\rm NB}$; i.e., the mean redshift of the selected NB) and the true redshift of the mock object ($z_{\rm spec}$) is smaller than the width of a NB in redshift space ($\Delta z\lesssim 0.12$; see Table~\ref{tab:NB_lya_redshifts}). We train ANN and random forest (RF) regressors with the aim of retrieving the spectroscopic redshift for the objects in our test set. The input for the ML redshift regressors is the same as for the ANN classifier described in Sect.~\ref{sec:ML_class}.
We perform a search of optimal parameters for the ANN and RF algorithms over an wide grid. We find that the RF regressor greatly outperforms the ANN and is less affected by over-fitting for the optimal parameters of both.

Bottom panel of Figure~\ref{fig:redshift_phot_spec} shows the difference between the redshift obtained by the RF and the spectroscopic redshift of all the selected objects in the PAUS catalogs with spectroscopic counterparts in SDSS, DESI and HETDEX. The initial NB redshifts are biased with respect to those of the spectroscopic counterparts, and present larger variance than the redshifts obtained by the RF: the bias and standard deviation are $(\mu,\sigma)(\Delta z)=(0.04, 0.03)$ for the initial NB redshifts, and $(\mu,\sigma)(\Delta z)=(0.003, 0.013)$ and $(0.010, 0.012)$ for the redshifts produced by the RF, for the SDSS and DESI spectroscopic sub-samples, respectively. The redshift bias of the DESI spectroscopic sub-sample is slightly larger than that of the SDSS sample. This can be explained by the fact that the mocks employed to train the RF algorithm are based on SDSS spectra. However, the RF estimation significantly improves the precision of the redshift measurements within the global spectroscopic sample.

\subsubsection{\lya luminosity correction}

The luminosity of the selected \lya lines is estimated as
\begin{equation}
    \Llya \approx \left(f_{\rm NB}^\lambda - f_{\rm cont}^\lambda\right) \cdot {\rm FWHM}_{\rm NB}\cdot 4\pi d_L^2(z)\,, \label{eq:L_lya}
\end{equation}
where $f_{\rm NB}^\lambda$ and $f_{\rm cont}$ are the NB and continuum fluxes; FWHM$_{\rm NB}$ the full width at half maximum of the NB filter, and $d_L$ the luminosity distance according to our cosmology. Several effects can alter the measurement of the \lya flux. In the first place, the estimation of $f_{\rm cont}$ is performed assuming a mean IGM transmission for the filters at shorter wavelengths than \lya (see Sect.~\ref{sec:selection_method}), yet there can be significant variability in the \lya forest for individual sources, especially at the lowest redshift bins \citep[e.g.,][]{Rollinde13}. Emission lines from QSOs typically show broad profiles, so the tails of the \lya profile may be partially out of the coverage of a given NB. This effect is also enhanced if the intrinsic \lya peak is not well aligned with the NB effective wavelength. In addition, in Eq.~\ref{eq:L_lya} a top-hat filter profile is assumed \footnote{Nonetheless, this is a reasonably good approximation for the PAUS NBs, and its effect should be minimal.}. Finally, while most spectroscopic surveys are capable of disentangling the \ion{N}{V} QSO line ($\lambda 1240.81$ \AA) from \lya , this task is not possible with NB photometry, since the \lya and \ion{N}{V} are cannot be separated within the single flux measurements obtained with PAUS NBs. The flux of \ion{N}{V} is typically fainter than \lya, but it can significantly affect the \lya measurement \citep[see e.g.,][]{Vanden-Berk01, Selsing16}.

We empirically correct for these biases based on the comparison of our measured \lya fluxes with the spectroscopic determinations of the SDSS DR16 counterpart sub-sample. For this, we determine the systematic offset between our flux measurements and those of SDSS as a function of $\Llya$, and subtract it from the flux measurements of our general sample (up to $\Delta\logten\Llya\sim 0.2$; see Fig.~4 in \citetalias{Torralba-Torregrosa23}).

\subsection{Purity and completeness}\label{sec:puricomp}

Following \citetalias{Torralba-Torregrosa23}, we compute 2D maps of purity and number count corrections in order to estimate the \lya LF. We apply our selection method (Sect.~\ref{sec:selection_method}) to the mocks, and distribute the selected mock sample over a 2D grid of measured \lya luminosity and synthetic $r$ band magnitude (i.e., $\Llya^{\rm obs}$, $r_{\rm synth}$). We then compute the 2D purity as
\begin{equation}
    % P^{\rm 2D} = \frac{\mathrm{TP}}{\mathrm{TP} + \mathrm{FP}}\,,
    P^{\rm 2D} = \frac{N_{\rm sel.,\,true}}{N_{\rm sel.}}
    \label{eq:purity}
\end{equation}
and the number count correction as
\begin{equation}
    w^{\rm 2D} = \frac{N_{\rm tot.}}{N_{\rm sel.,\,true}}
    \label{eq:completeness}
\end{equation}
where $N_{\rm sel.}$ is the number of selected objects, $N_{sel.,\,true}$ the number of correctly selected objects (i.e., \lya emitting quasars at the correct redshift), and $N_{\rm tot.}$ the total number of objects in the mock, which would all be selected if the method's completeness was 100\%. This quantities are evaluated in each cell of the 2D grid.
These estimates of $P^{\rm 2D}$ and $w^{\rm 2D}$ implicitly include the purity and completeness components of the NB flux excess selection and the ANN classifier. The visual inspection of our selected sample was performed only to remove artifacts unforeseen by the mocks, and thus we assume it does not affect the completeness.
We also neglect the incompleteness arising from the intrinsic limitations of the reference catalogs, as these catalogs are based on significantly deeper data and are complete up to $i\sim 24.5$ (see Sect.~\ref{sec:observations}).

\begin{figure}
    \centering
    \includegraphics[width=\linewidth]{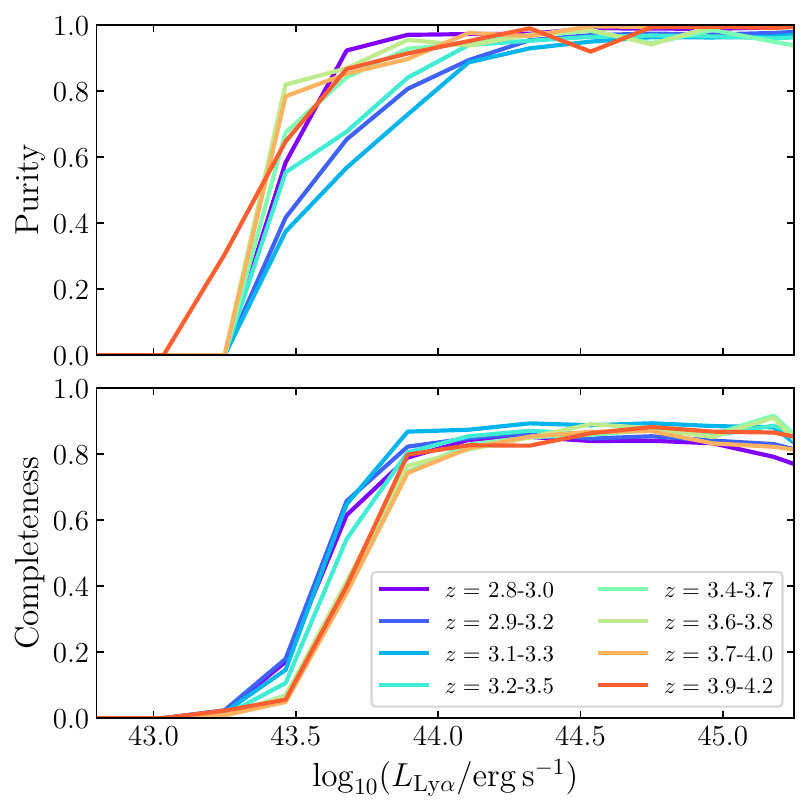}
    \caption{1D purity and completeness of our sample as a function of measured $\Llya$. We show this values in the bins of redshift used to estimate the \lya and UV LFs (see Sect.~\ref{sec:lya_LF_computation}). These curves were estimated by applying the selection method to the mocks, in bins of measured \lya luminosity. The purity of the sample increases with $\Llya$, reaching values of $\sim 1$ for the high-end at all redshifts. The completeness, on the other hand, increases monotonically with $\Llya$, reaching maximum values of $\approx 85\%$ at $\Llya \approx 10^{44}$ erg\,s$^{-1}$ at all redshifts.}
    \label{fig:1d_puricomp}
\end{figure}

By marginalizing the 2D purity and completeness maps over the $r_{\rm synth}$ axis we estimate the total purity and completeness of our sample in terms of $\Llya$. The 1D purity and completeness estimations from our mocks is shown in Fig.~\ref{fig:1d_puricomp}. Figure~\ref{fig:2D_purity_and_comp} is the 2D version of this figure in bins of $r$ and $\Llya$. The purity of the sample reaches $\sim 90\%$ at $\log_{10}(\Llya / {\rm erg\,s}^{-1})\approx 44$ for all the considered redshift bins. Conversely, the completeness reaches a maximum value of $\sim 85\%$. Interestingly, the maximum completeness values correspond to the higher redshift bins. Fig.~\ref{fig:1d_puricomp} also shows that the purity increases more rapidly with luminosity at the highest redshift bins. The reason for this is that the 7 NBs with lowest effective wavelengths are more likely to suffer from higher contamination from QSO \ion{C}{IV} lines; the rest of the NBs probe redshifts in which the \lya line is observed in combination with \ion{C}{IV}, allowing our method to identify the redshift correctly. This effect reflects in Fig.~\ref{fig:redshift_phot_spec} (top panel). This figure also illustrates that the purity of the sub-sample with spectroscopic counterparts is very close to unity for a large fraction of our $z_{\rm phot}$ range, especially for $z_{\rm phot}>3.25$, due to the decrease of line confusions.

\subsection{\lya luminosity function computation}\label{sec:lya_LF_computation}

In order to compute the \lya LF we follow the same scheme detailed in \citetalias{Torralba-Torregrosa23}, adapting those methods to the specific case of this work. In order to obtain the \lya LF, we need to characterize the purity and completeness of our LAE candidate sample, and apply the adequate corrections.
Each candidate $k$ is assigned a weight $w=w^{\rm 2D}$, being $w^{\rm 2D}$ the number count correction from Eq.~\ref{eq:completeness}. On the other hand, we introduce the parameter $p_k$ whose value is zero if the $k$-th candidate is rejected, and unity if it is accepted to be included in the LF computation. Each candidate is accepted for the LF computation with a probability equal to the value of the purity $P^{\rm 2D}_k$ (Eq.~\ref{eq:purity}).

The \lya LF for each considered NB set is computed an arbitrarily large number of times ($n=1000$) to achieve significant statistics. The $i$-th determination is computed as
\begin{equation}
    \Phi_i [\logten \Llya ] = \frac{\sum\limits_k p_k \cdot w_k}{V \cdot \Delta \left( \logten\Llya \right)} \,,\label{eq:Lya_LF_computation}
\end{equation}
where the index $k$ runs through the LAE candidates within the redshift interval defined by the given NB set, in bins of \lya luminosity; $V$ is the maximum volume our survey can probe \citep{Schmidt68}, determined by the covered interval of wavelengths by the NBs and the survey area\footnote{The covered volume $V$ is computed by integrating the Astropy function \texttt{differential\_comoving\_volume} over the redshift interval covered by each NB, and multiplying by the effective survey area.}. On each iteration, the \lya luminosity and $r$ flux is randomly perturbed using Gaussian uncertainties, and the parameters $p_k$ and $w_k$ are recomputed accordingly. In addition, a multiplicative factor $(1 - 0.08)^{-1}$ was applied to the LF in order to account the incompleteness due to the removal of sources with neighbours closer than 3\arcsec from the parent sample (see Sect.~\ref{sec:parent_sample}).

In order to include the effect of cosmic variance and field-to-field variation effects, the LF is computed for bootstrapped sub-regions of the wide PAUS fields. We define tiles of equal area ($\sim 0.12$ deg$^2$) for each of the three PAUS wide fields. These tiles are obtained by generating random points, from an uniform distribution in spherical coordinates, within the PAUS masks and defining regions with equal number of random points inside. Finally, the LF is obtained as the median of all the $\Phi_i$ determinations obtained from the bootstrapped regions. The final errors on our LF are computed as the 16th and 84th percentiles of the same $\Phi_i$ distribution. Finally, the UV LF can be computed in a similar way, using Eq.~\ref{eq:Lya_LF_computation} substituting $\logten\Llya$ with the absolute UV magnitude computed at rest-frame 1450 \AA, $M_{\rm UV}$.

%%%%%%%%%%%%%%%%%%%%%%%%%%%%%%%%%%%%%%%%%%%%%%%%%%%%%%%%%%%%%%%%%%%%%%%%%%%%%%%%%%%%%%%%%%%%%%%%%%%%%%%%%%%%%%%%%%%%%%%%%%%%%%%%%%%%%%%%
\section{Results}\label{sec:results_and_discussion}

In this Section, we first present the composite QSO spectrum, obtained from the stack of NB fluxes of the objects in our sample (Sect.~\ref{sec:QSO_composite_spectrum}). Next, we present the \lya and UV luminosity functions estimated from our data (Sects.~\ref{sec:multi_LF} and \ref{sec:UV_LF}).

\subsection{QSO composite spectrum}\label{sec:QSO_composite_spectrum}

\begin{figure}
    \centering
    \includegraphics[width=\linewidth]{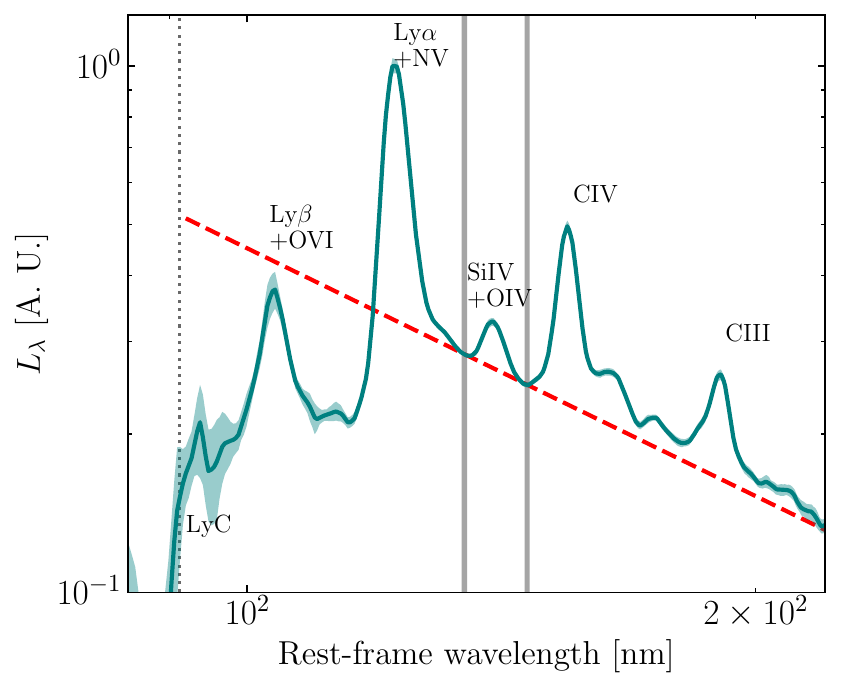}
    \caption{Stacked composite spectrum of all QSOs in our golden sample. We show the mean composite spectrum (teal solid line) and bootstrap 1$\sigma$ interval (shaded region). The shaded grey regions mark the rest-frame wavelength intervals used to fit a power-law continuum. We represent the fitted power-law as a dashed red line, whose slope is $\beta_{\rm UV}=-1.55\pm 0.02$. We show the labels of the most important QSO emission lines and the LyC limit (91.2 nm).}
    \label{fig:stacked_QSO_all}
\end{figure}

We stack the NB SEDs of our golden sample (see Sect.~\ref{sec:visual_inspection}) in order to obtain a photometric composite QSO spectrum. For this task we use the code \texttt{stonp}\footnote{https://github.com/PAU-survey/stonp}, developed to stack multi-band photospectra in a common rest-frame wavelength grid. Before stacking, each spectrum is normalized dividing by the $r_{\rm synth}$ band (see Sect.~\ref{sec:ML_class}). The composite spectrum is shown in Fig.~\ref{fig:stacked_QSO_all}, and it reproduces the expected features of a QSO spectrum: power-law continuum with a break caused by the \lya forest ($\lambda_0<121.56$ nm), and total absorption past the Lyman limit ($\lambda_0<91.2$ nm); and the typical Ly$\beta$+\ion{O}{VI}, \lya+\ion{N}{V}, \ion{Si}{IV}+\ion{O}{IV}, \ion{C}{IV} and \ion{C}{III} broad emission lines.

We fit a power-law to the QSO UV continuum, using the rest-frame wavelength windows $134<(\lambda_0/{\rm nm})<135$ \& $146<(\lambda_0/{\rm nm})<147$ (Fig.~\ref{fig:stacked_QSO_all}).
The choice of these windows is motivated by the goal of capturing the power-law shape of the spectrum in rest-frame wavelength intervals that do not contain emission lines nor any prominent non-continuum emission like the \ion{Fe}{II} emission complex \citep[see, e.g.,][]{Vanden-Berk01}.
The spectral index (i.e., $1+\beta_{\rm UV}$, being $\beta_{\rm UV}$ the continuum power-law slope) of QSOs is known to vary with the choice of the rest-frame wavelength window, as its measurement is susceptible to by affected broad-emission features \citep[e.g., thermal emission from the accretion disk, \ion{Fe}{II} emission series etc.; see e.g.,][]{Shull12}. For this fit, which will serve as reference, we use the stack of every QSO in the full redshift range in order to maximize the S/N. We obtain a best-fitting UV slope of $\beta_{\rm UV}=-1.55 \pm 0.02$ for the global stack ($f^\lambda \propto \lambda^{\beta_{\rm UV}}$). This result is in line with the slope measured by \cite{Vanden-Berk01} for a SDSS composite spectrum using a fitting window of $135<(\lambda_0/{\rm nm})<136.5$ ($\beta_{\rm UV}=-1.56$). More recently, \cite{Lusso15} measured a UV slope of $\beta_{\rm UV}=-1.61\pm0.01$, using a different choice for the wavelength windows (109.5--111.0, 113.5--115.0, 145.0--147.0, 197.5--201.0, and 215.0--220.0 nm).

\subsection{The \lya luminosity function at $2.7\lesssim z\lesssim5.3$}\label{sec:multi_LF}

\begin{table}
    \centering
    \caption{Intervals of NBs used for the \lya and UV LF computation.}
    \begin{tabular}{llccc}
    \toprule
    NB$_{\rm min}$ & NB$_{\rm max}$ & $z_{\rm min}$ & $z_{\rm max}$ & Volume [$10^6$ Mpc$^{3}$]\\
    \midrule
    0 & 3 & 2.71 & 3.04 & 4.167\\
    2 & 5 & 2.86 & 3.21 & 4.089\\
    4 & 7 & 3.03 & 3.38 & 4.090\\
    6 & 9 & 3.20 & 3.55 & 4.044\\
    8 & 11 & 3.35 & 3.71 & 3.975\\
    10 & 13 & 3.52 & 3.87 & 3.855\\
    12 & 15 & 3.68 & 4.03 & 3.799\\
    14 & 18 & 3.85 & 4.28 & 4.634\\
    19 & 30 & 4.26 & 5.26 & 10.481\\
    \bottomrule
    \end{tabular}
    \tablefoot{The NB indices shown in the first two columns refer to the NB IDs defined in Table~\ref{tab:NB_lya_redshifts}.}
    \label{tab:LF_NB_intervals}
\end{table}

\begin{figure*}
    \centering
    \includegraphics[width=\linewidth]{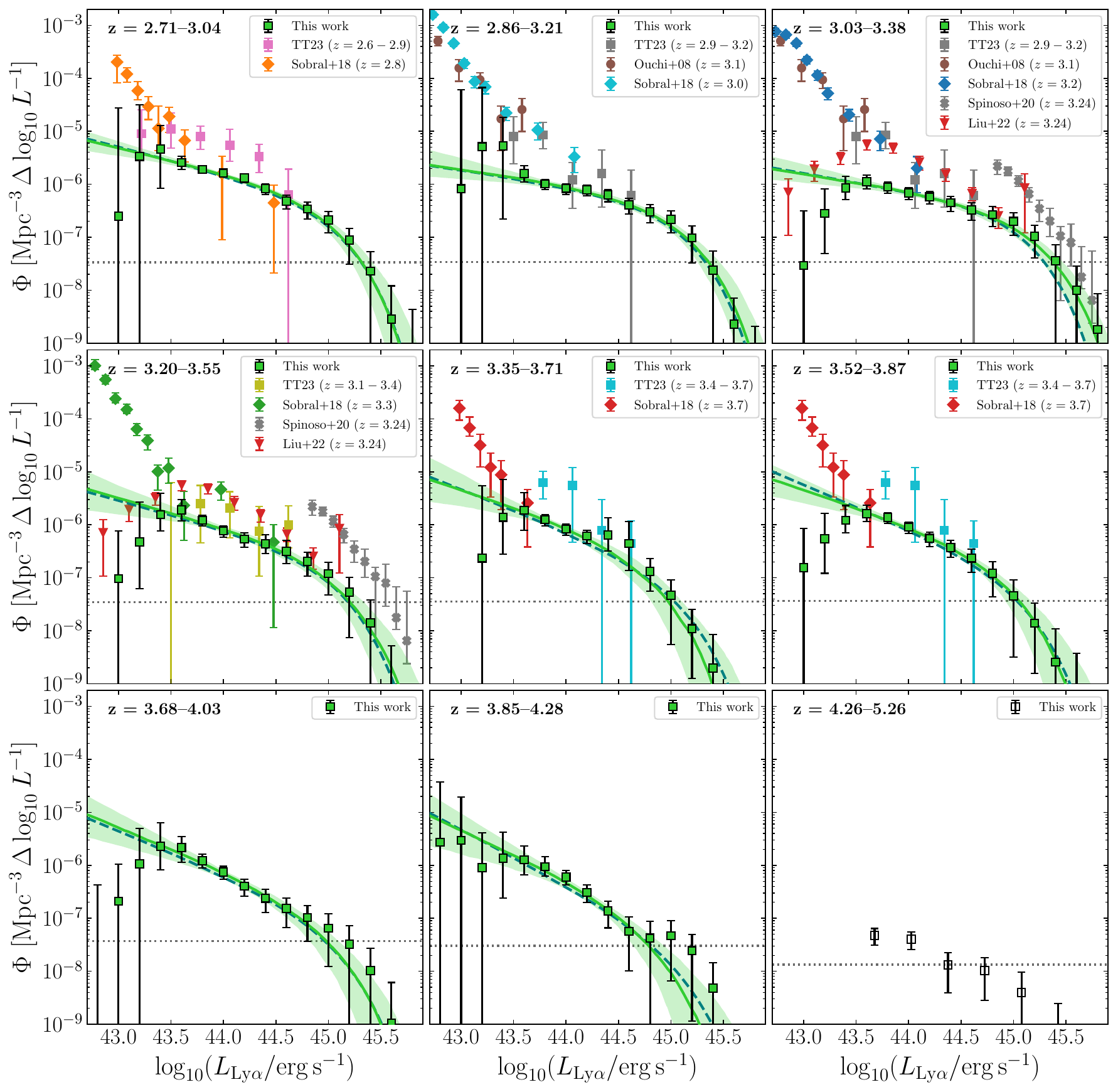}
    \caption{\lya luminosity function in 9 intervals of redshift defined in Table~\ref{tab:LF_NB_intervals}. We compare with \lya LFs in the literature at similar redshifts. The green solid line and blue shaded area show the best Schechter fit with free parameters and the $1\sigma$ confidence interval, respectively. The dashed blue line shows the Schechter best-fit with for fixed $L^*=45$ erg\,s$^{-1}$. The \lya LF of the highest redshift bin shown in this figure was computed assuming purity and completeness equal to 1, hence it should be regarded as a lower limit (See Sect.~\ref{sec:multi_LF}). The horizontal gray dotted line marks the 1 object per bin level. Data-points below this limit signal LF bins which contain less than 1 object after we average-out the 1000 LF determinations used to compute our final LF.}
    \label{fig:multi_LF}
\end{figure*}

Figure~\ref{fig:multi_LF} shows the \lya LF in the 9 overlapping intervals of redshift as defined in Table~\ref{tab:LF_NB_intervals}. We discuss the evolution of our LF in Sect.~\ref{sec:Lya_LF_discussion} below. The highest redshift bin in Fig.~\ref{fig:multi_LF} is computed from the high-$z$ sample (see Sect.~\ref{sec:high_z_sample}), and assuming purity and completeness equal to 1. The reason for this is that our 2D purity and number-count correction estimations cannot be accurately computed for $z\gtrsim 4.5$, where the SDSS DR16 sample begins to be scarce. As discussed in Appendix C of \citetalias{Torralba-Torregrosa23}, purity inaccuracies do not significantly affect the estimation of the LF, and completeness smaller than 1 induces a positive correcting factor in the LF. Hence, the results of this highest-$z$ bin should be taken as lower limits to the \lya LF. We also note that, due to the limited number of candidates for such high redshifts, this last bin is considerably wider than the rest, covering $z\approx4.26$--$5.26$.

\subsubsection{Schechter fit to our data}\label{sec:lya_sch_fit}

We fit a Schechter function \citep{Schechter76} to our data, of the form
\begin{equation}
    \Phi [L] {\rm d}(\logten L) = \log 10 \cdot \Phi^* \cdot \left ( \frac{L}{L^*}\right)^{\alpha + 1} \cdot {\rm e}^{-L / L^*} {\rm d}(\logten L)\,.
\end{equation}\label{eq:schechter}
We perform the fitting using the Monte-Carlo Markov Chain (MCMC) technique, aided by the python package \texttt{autoemcee}\footnote{https://github.com/JohannesBuchner/autoemcee}. Fit our \lya LF in the full measured range for consistency. In all panels of Fig.~\ref{fig:multi_LF}, a deviation from the characteristic power-law shape of the Schechter function is evident at the faint end of the LFs. This effect is likely due to inaccuracies in the completeness and purity models at such low luminosity intervals. However, our method effectively accounts for these inaccuracies by generating large uncertainties, which reduce the influence of these bins on the fitted curve.

The Schechter fit is performed using a normal log-likelihood and wide flat priors for all three free parameters: $\logten(\Phi^* / {\rm Mpc^{-3}}) \in [-9, -3]$, $\logten (L^* / {\rm erg\,s^{-1}}) \in [43, 46]$ and $\alpha \in [-3, -1]$. The best-fit parameters are summarized in Table~\ref{tab:sch_params}. The best-fitting value for the normalization $\Phi^*$ marginally shows anti-correlation with redshift. The effect of this evolution can be easily seen as the overall decline towards higher redshift of the \lya LF in Fig.~\ref{fig:multi_LF}. The faint-end power-law slope $\alpha$ also shows steepening values with redshift, while the characteristic luminosity $L^*$ has no clear evolutionary trend in the probed redshift range.

We perform further Schechter fits  to our \lya LFs fixing $\logten L^*=45$ erg\,s$^{-1}$. Fixing this parameter removes the possible correlations with $\Phi^*$ ($\Phi^*$ defines the value of the LF at $L=L^*$, see Eq.~\ref{eq:schechter}). The evolution of the parameter $\Phi^*$ becomes more evident, and the faint-end slope shows a very similar steepening trend with redshift. We discuss the fitted parameters more in detail in Sect.~\ref{sec:Lya_LF_discussion}.

\begin{figure*}
    \centering
    \includegraphics[width=\linewidth]{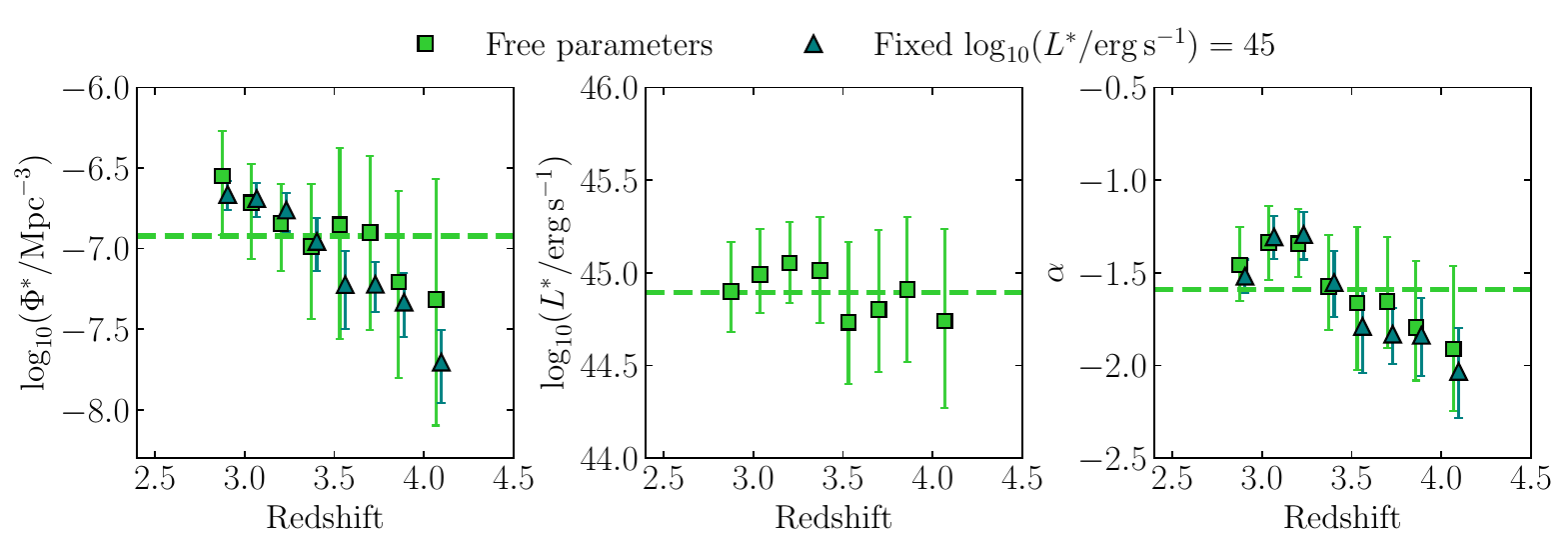}
    \caption{Best-fit Schechter parameters for the LFs in the 9 intervals of redshift. The horizontal dashed lines represent the average best-fit Schechter parameters across all redshifts. When leaving all three parameters free (green squares), the normalization $\Phi^*$ and the characteristic luminosity $L^*$ show a slight decline within the intervals of confidence, whilst the faint-end slope $\alpha$ shows no clear trend. We also show the fitted parameters when fixing $\alpha$ and $L^*$ to the average values marked with the green dashed lines (blue triangles). The values of the latter are slightly shifted in the horizontal axis for visual clarity.}
    \label{fig:schechter_parameters}
\end{figure*}

\subsubsection{Integrated \lya luminosity function}\label{sec:integrated_lya_LF}

\begin{figure}
    \centering
    \includegraphics[width=\linewidth]{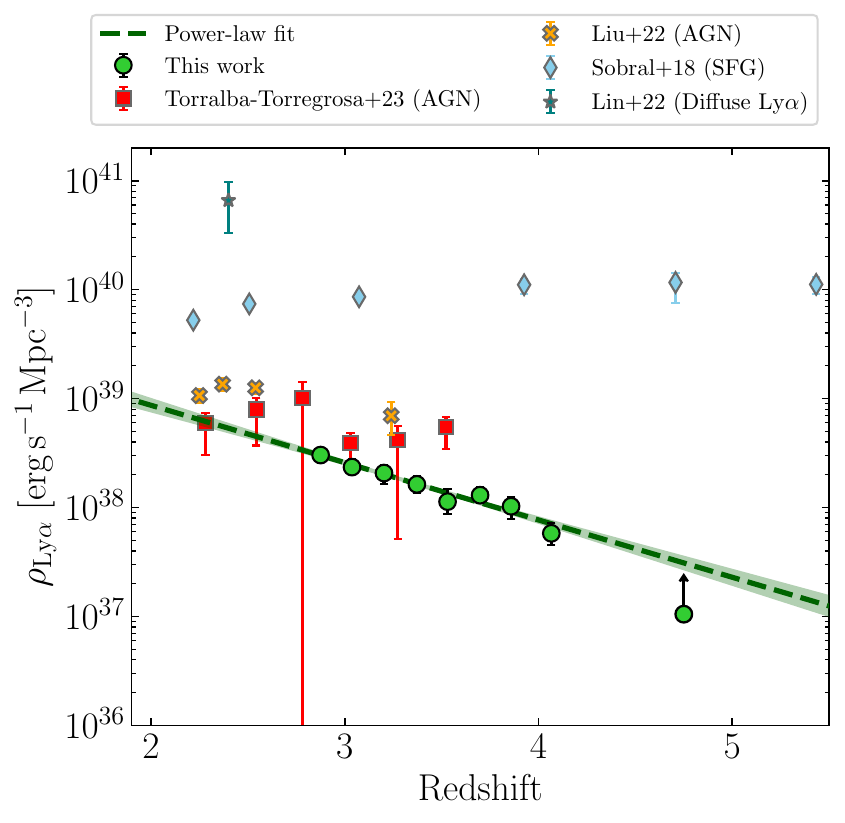}
    \caption{\lya luminosity density per unit comoving volume. These points were obtained by integrating the \lya LF for $\Llya > 10^{43.5}$ erg\,s$^{-1}$. Our results show a declining trend with increasing redshifts, compatible with the results presented in \citetalias{Torralba-Torregrosa23} and with the integrated \lya LF from \protect\cite{Liu22b}. We also compare with the integrated LF of the SFG population in \protect\cite{Sobral18}, which covers $\Llya \lesssim 10^{43.5}$ erg\,s$^{-1}$, showing that the integrated AGN \lya luminosity is increasingly smaller than the galaxy contribution at higher redshift. We also compare with the diffuse \lya emission measurement of \protect\cite{Lin22}.}
    \label{fig:integrated_Lya_density}
\end{figure}

In Fig.~\ref{fig:integrated_Lya_density}, we show the results of integrating the best-fit of the \lya luminosity function in the range $\log_{10}(L_\text{\lya} / {\rm erg\,s}^{-1})> 43.5$, for the redshift intervals shown in Fig.~\ref{fig:multi_LF}. The resulting total volume density of \lya emission $\rho_\text{\lya}$ shows a clear declining trend towards increasing redshifts. We fit this trend to $\log_{10}(\rho_\text{\lya} / {\rm erg\,s}^{-1}{\rm\,Mpc}^{-3})=b + a\cdot z$, obtaining the parameters $a=-0.53 \pm 0.05$ and $b=39.90\pm 0.16$. For this fit, we excluded the highest redshift bin (empty circle in Fig.~\ref{fig:integrated_Lya_density}, however this lower-limit is in agreement with the extrapolation of the fitted evolution of $\rho_\text{\lya}$.

Our results are compatible with the measurements obtained by integrating the best-fits of the data presented in \citetalias{Torralba-Torregrosa23}, for most of the redshift bins. The surveyed area employed in \citetalias{Torralba-Torregrosa23} was significantly smaller (by a factor $\sim 30$), and the discrepancies can be due to cosmic variance. The extrapolation of our $\rho_\text{\lya}$ best-fit is considerably lower than the result of integrating the double power-law fits in \cite{Liu22b}, in their $z=2.25, 2.37$ and $2.54$ bins, in the same luminosity range. These differences are also reflected in the discrepancies with their faint-end of the \lya LF that can be seen in Fig.~\ref{fig:multi_LF}. In their work, they used blind spectroscopy to select a sample of much fainter AGNs, which also include narrow-line AGNs and sources with significant contribution from star formation in the host galaxy, while our sample is mainly composed of luminous broad-line quasars. 

We also compare the presented measurements of $\rho_\text{\lya}$ for the AGN \lya LF with the measurements of \cite{Sobral18}, for the faint-end of the \lya LF, populated by star-forming galaxies. The value of $\rho_\text{\lya}$ for this population increases with redshift, showing the opposite trend to the bright \lya emitting AGN population. Our results show that for $z\approx 4$, the total emissivity of AGN \lya is subdominant by a factor of $\approx 200$, with respect to the star-forming LAEs. Moreover, we also display the estimated $\rho_\text{\lya}$ from unresolved emitters and diffuse IGM emission from \cite{Lin22}; this value is expected to be significantly larger than than \lya luminosity from resolved emitters \citep{Byrohl2022}, but can be explained by integration of faint LAE LFs \citep[e.g.,][]{Drake17, Sobral18} to an arbitrarily faint end \citep{Renard2024}.

\subsection{UV luminosity function}\label{sec:UV_LF}

\begin{figure*}
    \centering
    \includegraphics[width=\linewidth]{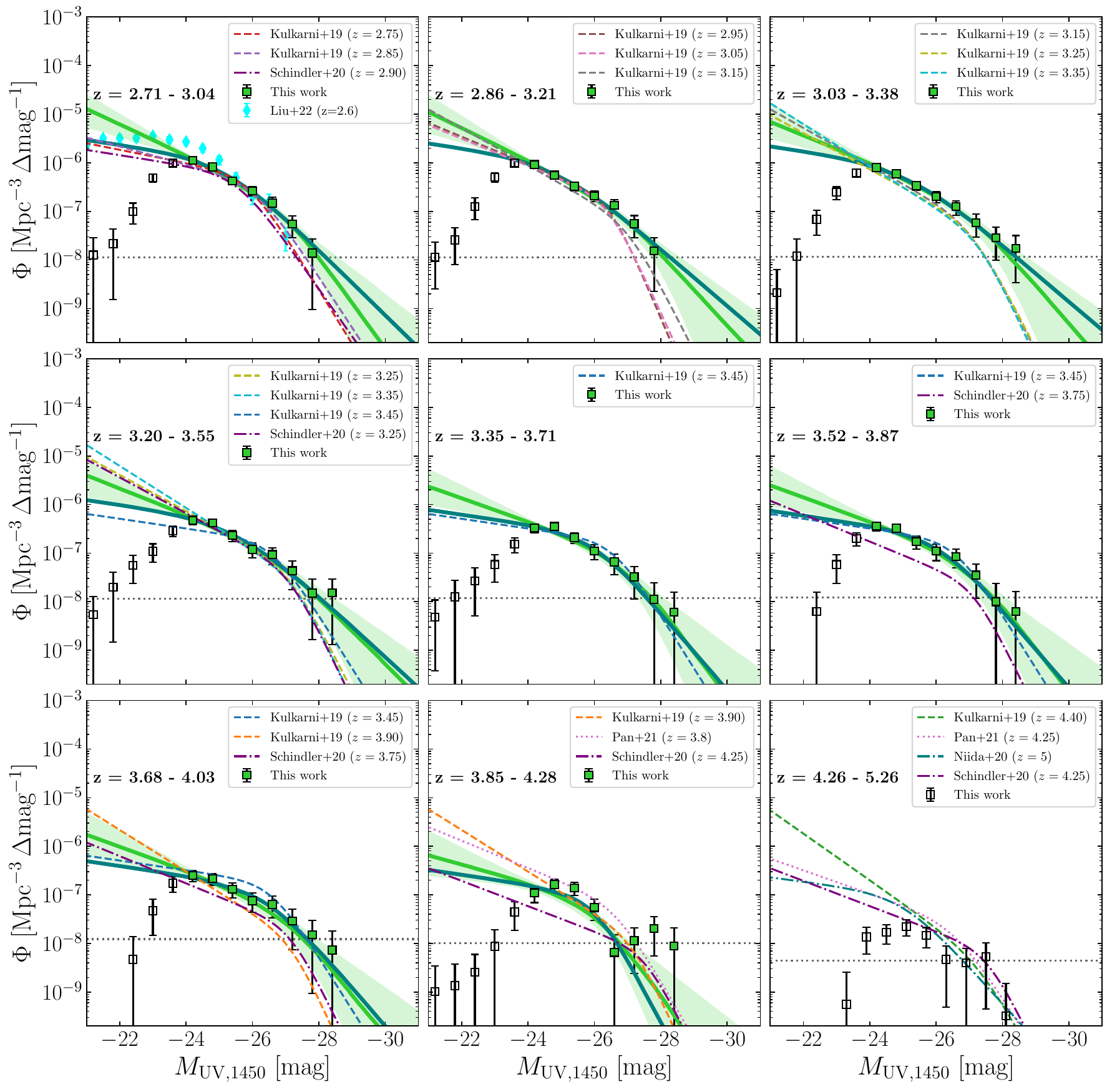}
    \caption{UV luminosity function of our sample selected via \lya emission. We show the LF in the same redshift bins of the \lya LF (Fig.~\ref{fig:multi_LF}). The filled green squares are used to fit a double power law, while the empty squares are discarded in the fit, due to low completeness (average completeness $<50\%$). The solid green line and shaded area show the best-fit and 1$\sigma$ uncertainties to a double power-law (teal solid line shows the fit with fixed $\beta$). We compare with double power-law fits in the literature, for the redshifts that approximately fall in our bins. The UV LF of the highest redshift bin shown in this figure was computed assuming completeness equal to 1, hence it should be regarded as a lower limit (See also Fig.~\ref{fig:multi_LF} and Sect.~\ref{sec:multi_LF}). We mark the 1 object per bin limit as a horizontal dotted line. Data points below this limit signal LF bins which contain less than 1 object after we take the average of the 1000 LF determinations used to compute our final LF.}
    \label{fig:multi_UV_LF}
\end{figure*}

One of the advantages of our multi-NB data is the possibility to measure SEDs for all the objects in the field without pre-selection, in an extensive range of wavelengths. We measure the UV absolute magnitude ($M_{\rm UV}$) for every object in our selected sample, defined as the luminosity at rest-frame wavelength 145 nm \citep[e.g.,][]{Glikman10}. The UV magnitude is thus computed as
\begin{equation}
    M_{\rm UV} = m_{\rm NB} - {\rm DM}(z) - K\,,
\end{equation}
where $m_{\rm NB}$ is the apparent magnitude of the NB corresponding to the rest-frame wavelength 145 nm of each object, DM$(z)$ the redshift dependent distance modulus according to our cosmology, and $K$ the $K$-correction defined as
\begin{equation}
    K = -2.5\log_{10}(1 + z)^{\alpha + 1}\,,
\end{equation}
where $\alpha$ is the spectral index \citep[i.e., $\alpha = -2-\beta_{\rm UV}$; see e.g.,][]{Wisotzski00}. For the computation of the $K$-correction we adopt the UV slope measured in our NB composite spectrum ($\beta_{\rm UV}=-1.55$, $\alpha=-0.45$; see Sect.~\ref{sec:QSO_composite_spectrum}). We note that the continuous coverage of the PAUS filters allow us to photometrically measure $M_{\rm UV}$ in the adequate window without having to apply bandpass corrections.

We obtain the UV LFs for our sample following a very similar procedure to that used for obtaining the \lya LF, described in Sect.~\ref{sec:lya_LF_computation}. The median purity and number count corrections based on $r_{\rm synth}$--$\Llya$ are applied, because the selection of the sources which make part of the UV LF is still based on the same selection function as the \lya LF. Hence, our results presented in Fig.~\ref{fig:multi_UV_LF}, correspond to the UV LFs of our \lya-selected AGN sample. We compare our results with the best-fitting curves in the literature for similar redshift intervals \citep{Kulkarni19, Niida20, Schindler19, Pan22}.

The last redshift bin of our UV LF ($4.26<z<5.26$) is computed in the same way as for the \lya LF (see Sect.~\ref{sec:multi_LF}), using visually selected objects from a pre-selection by our method (see Sect.~\ref{sec:high_z_sample}). As for the case of the \lya LF (see introduction of Sect. \ref{sec:multi_LF}), the purity and completeness of this redshift bin is not well determined, thus this result may be taken as a lower limit for the UV LF. Nevertheless, the UV LF in this last redshift bin is in good agreement with $z\sim 5$ determinations in the literature \citep{Kulkarni19, Niida20, Pan22}.

\subsubsection{Double power law fit}\label{sec:dpl_fit}

The faint-end of our UV LF ($M_{\rm UV} \gtrsim -24$) is affected by the poor purity and completeness (see Fig.~\ref{fig:1d_puricomp}). For $M_{\rm UV} \lesssim -24$, the UV LF is typically described by a double-power law (DPL):
\begin{equation}
    \Phi\left[M\right] = \frac{{\Phi^*}}{10^{-0.4 (M^* - M)(1-\beta)} + 10^{-0.4 (M^* - M)(1-\gamma)}}\,. \label{eq:dpl_fit}
\end{equation}
We fit our UV LF using Eq.~\ref{eq:dpl_fit}, using flat priors: $\logten(\Phi^* / {\rm Mpc^{-1}}) \in [-10, -5]$, $M^* \in [-28, -22]$, $\beta \in [-3, -1]$, $\gamma \in [-6, -1]$. We obtain the best-fit parameters shown in Fig.~\ref{fig:DPL_parameters}. The fitted parameters are discussed in Sect.~\ref{sec:UV_LF_discussion} below.

\begin{figure}
    \centering
    \includegraphics[width=\linewidth]{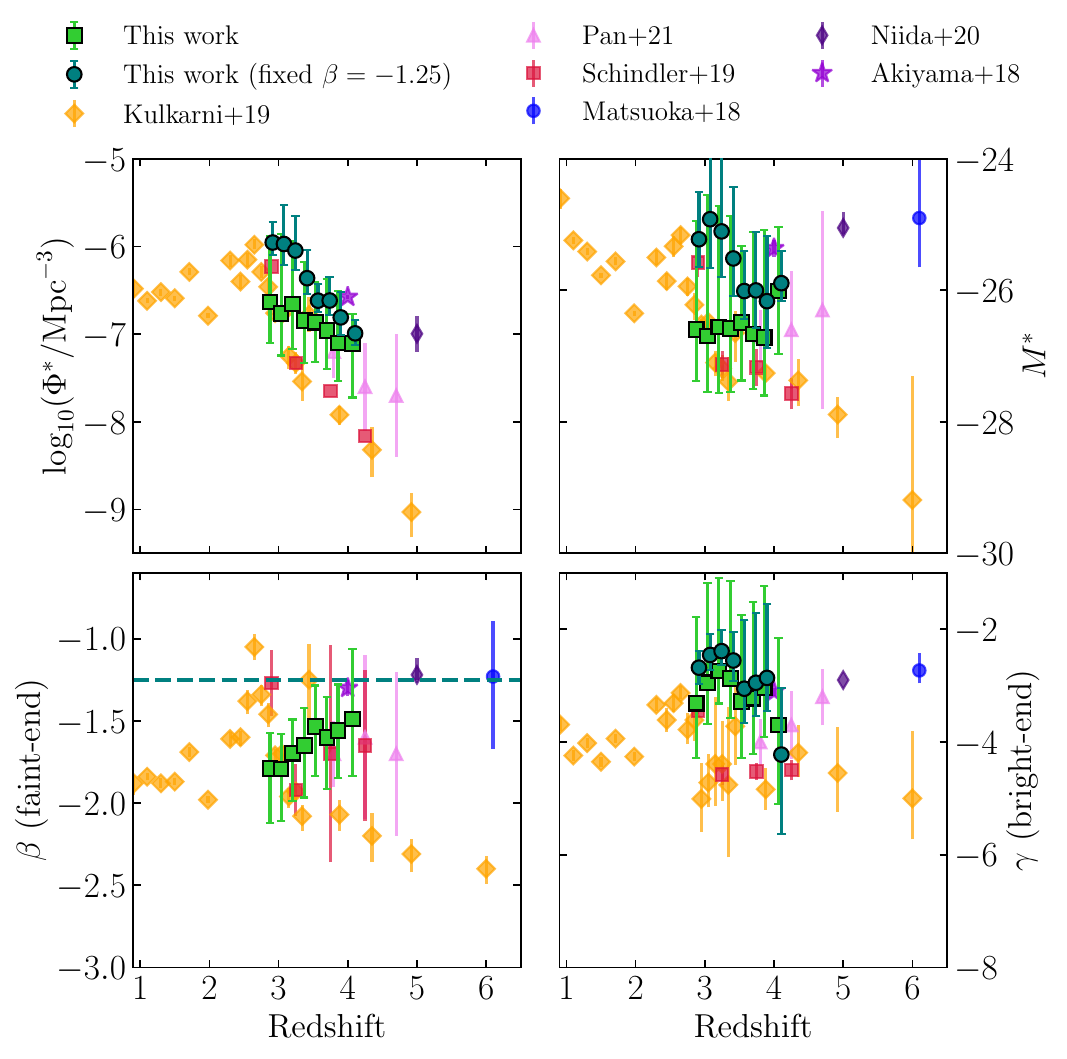}
    \caption{Best-fit double power-law parameters for the UV LF in bins of redshift. We show the results obtained by leaving all four parameters free (green squares) and fixing the faint-end slope ($\beta=-1.25$, shown as a horizontal dashed blue line; blue circles). We compare our results with those of \protect\cite{Kulkarni19}. Our results show a clear decrease in the value of the normalization parameter $\Phi^*$ with redshift, in line with the fitted Schechter parameter for the \lya LF. The faint-end slope $\beta$ also shows hints of a declining evolution, while the bright-end slope $\gamma$ is poorly constrained by our data.}
    \label{fig:DPL_parameters}
\end{figure}

\subsubsection{Integrated UV emissivity}\label{sec:eion_1450}

\begin{figure}
    \centering
    \includegraphics[width=\linewidth]{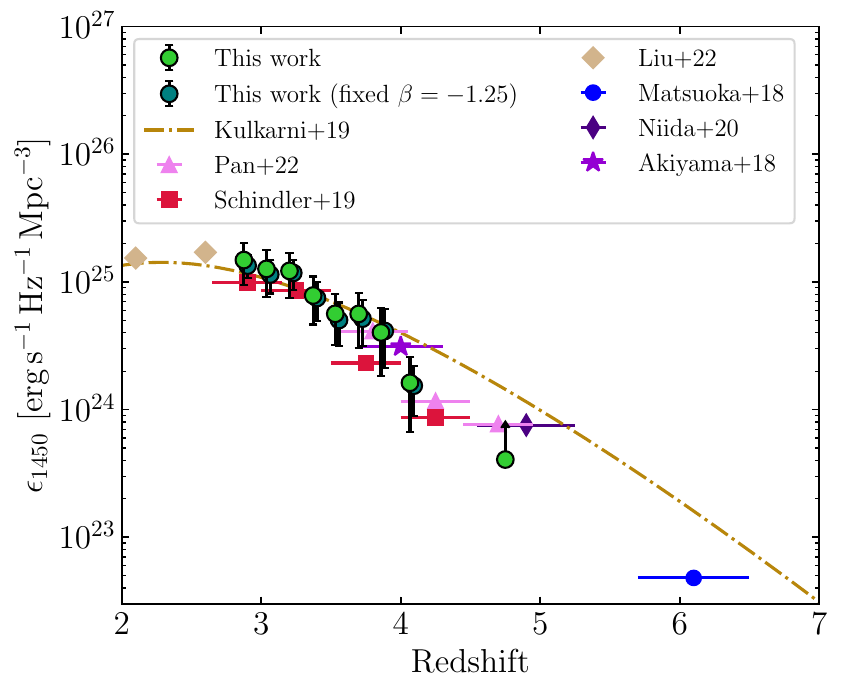}
    \caption{UV emissivity at rest-frame wavelength 1450 \AA . We show the emissivity obtained from the integral of the double power-law fit with four free parameters (green circles) and fixed faint end slope -1.25 (blue circles) -- the latter are slightly shifted in the horizontal axis for clarity. The total emissivity at 145 nm rapidly declines with redshift. This quantity is directly related to the ionizing photon emissivity (i.e., $\lambda_0 < 91.2$ nm), given that the AGN continuum is well approximated by a power-law. We compare our results with the integral of other UV LFs in the literature for $z>2$ \protect\citep{Matsuoka18, Akiyama18, Schindler19, Kulkarni19, Niida20, Pan22, Liu22b}.}
    \label{fig:eion_1450}
\end{figure}

The integral of the UV LF directly yields the total UV emissivity of the AGN population per unit volume. We integrate our best fits to obtain $\epsilon_{\rm 1450}$, the emissivity at 145 nm (i.e., the wavelength where $M_{\rm UV}$ is measured), by extrapolating from $M_{\rm UV}=-21$ to an arbitrarily luminous magnitude. This allows for a direct comparison with Fig.~8 in \cite{Kulkarni19}\footnote{In \cite{Kulkarni19}, the 91.2 nm emissivity is shown instead. This value is obtained extrapolating an assumed power-law continuum. We choose to directly show the 145 nm emissivity, re-scaling the values shown in their Fig.~8.}, as we show in our Fig.~\ref{fig:eion_1450}. Our Results are compatible with the fitted trend in \cite{Kulkarni19} as well as with the integrated UV LFs of several past works in the literature. We computed $\epsilon_{1450}$ in the same range for our best fits with fixed faint end slope ($\beta=-1.25$), obtaining similar values. Therefore, a small change in the faint-end slope has little impact on the total emissivity of intermediate-luminosity quasars.

\subsection{Mean Lyman-$\alpha$ forest transmission}\label{sec:IGM_trans_measurement}

\begin{figure*}
    \sidecaption
    \includegraphics[width=12cm]{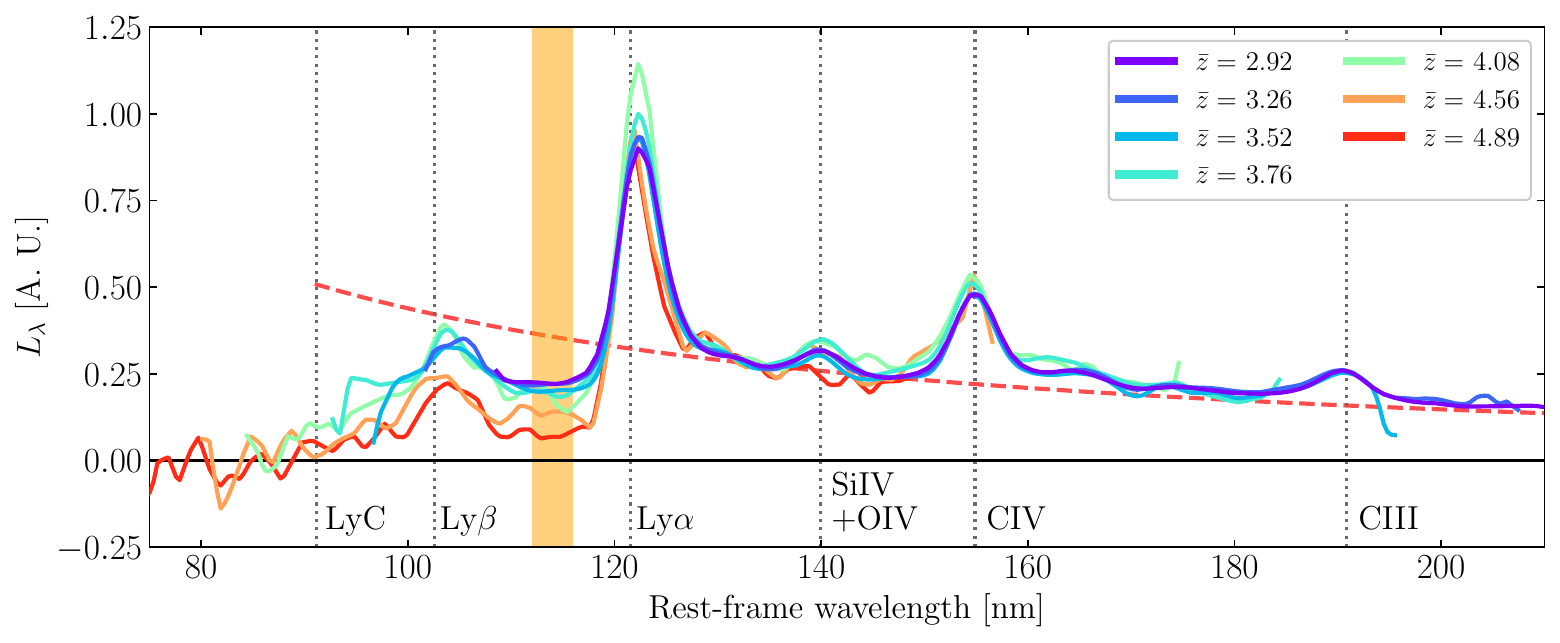}
    \caption{Composite QSO spectra in different bins of redshift. The composite spectra were obtained stacking the objects in our visually selected sample using the \texttt{stonp} code. The UV continuum slope show negligible evolution with redshift. The visible differences in the typical EW of the most prominent lines is likely due to intrinsically fainter sources dominating the lowest redshift bins, and the resolution increase in the rest-frame with redshift. The dotted vertical lines mark the rest-frame wavelengths of the most prominent QSO lines, and the Lyman limit ($91.2$ nm; LyC). The dashed pink line shows the best power-law fit for the UV continuum, with slope $\beta_{\rm UV}=-1.55 \pm 0.02$. The shaded orange region is used to compute the average IGM optical depth (see Sect.~\ref{sec:IGM_trans_measurement}).}
    \label{fig:composite_qso_spec}
\end{figure*}

\begin{figure}
    \centering
    \includegraphics[width=\linewidth]{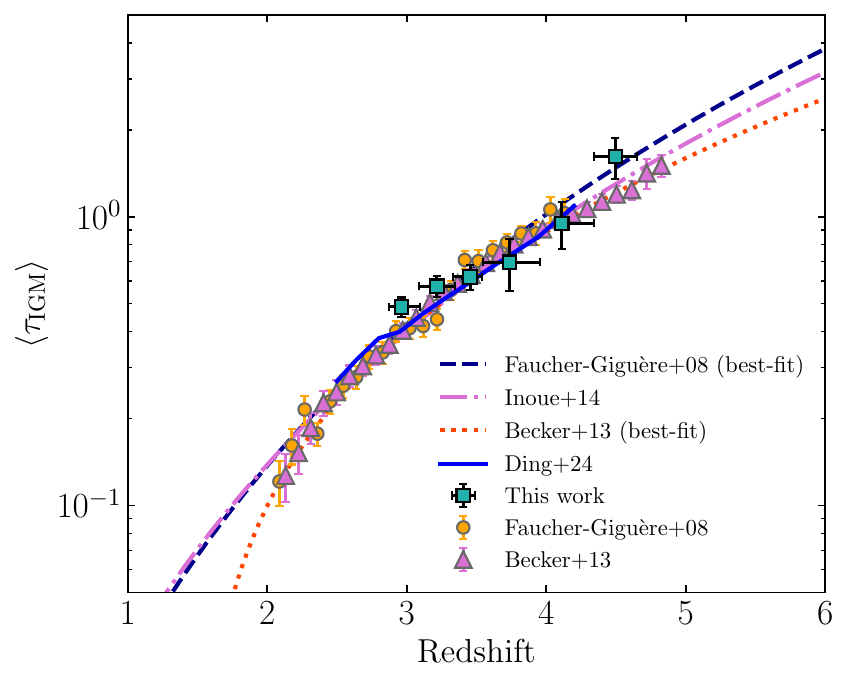}
    \caption{Mean IGM transmission due to the \lya forest. These estimations were obtained from the stacked composite spectra shown in Fig.~\ref{fig:composite_qso_spec}. As comparison, we show the estimations in \protect\cite{Becker13}, \protect\cite{Inoue14}, \protect\cite{ding2024} and \protect\cite{Faucher-Giguere08}, the latter employed as continuum correction in our selection method (See Sect.~\ref{sec:selection_method}).}
    \label{fig:IGM_lya_forest_T}
\end{figure}

In Sect.~\ref{sec:QSO_composite_spectrum} we stacked our sample in order to obtain a composite QSO spectrum in the full redshift range ($z=2.7$--5.3). Here we repeat the same procedure in bins of redshift. In Fig.~\ref{fig:composite_qso_spec} we show the stacked spectra in 8 bins of redshift. At wavelengths redward \lya, our QSO spectra are self-similar at all redshifts  --- the average UV spectrum of quasars is known to not significantly vary across cosmic time \citep[e.g.,][]{Onorato24}.
The EW of the most luminous emission lines (\lya and \ion{C}{IV}) is notably scattered, however this is probably due to fainter sources having higher S/N at lower redshifts, dominating the sample. The uncertainties in the systemic redshift can induce biases in the emission-line EWs of the stacked spectra. 
In addition, the spectral resolution of the NB pseudo-spectra increases with redshift, given that the rest-frame wavelength interval that is covered by a single NB is proportional to $\propto (1 + z)$. For this reason, the shape of the \lya line might also vary in the composite spectrum, once the latter is brought to the rest-frame (e.g., the \lya lines appear to be narrower in the two highest redshift bins).

On the other hand, the continuum in the rest-frame wavelength range within $91.2<(\lambda_0/{\rm nm})<121.57$ shows an evolution with redshift consistent with increasing mean IGM absorption by the \lya forest (+ Ly$\beta$ forest for $\lambda_0<102.5$ nm). 
We average the \lya forest transmitted flux of the composite spectra in each redshift bin, in the range $113<(\lambda_0/{\rm nm})<115$ (wavelength interval in which no prominent emission lines are expected; see Fig.~\ref{fig:composite_qso_spec}), and divide it by the extrapolated power-law flux fitted above, in order to obtain the mean IGM absorption due to the \lya forest. We show the mean IGM optical depth ($\tau_{\rm IGM}$) in Fig.~\ref{fig:IGM_lya_forest_T}, which relates to the IGM transmission as $T_{\rm IGM} = {\rm e}^{-\tau_{IGM}}$.
For this analysis, we only consider redshifts where the observed window used to estimate the transmitted flux is covered at least by the second NB in our set (NB465), to avoid border effects of the stack.
Our measurements are in line with previous determinations in the literature \citep{Faucher-Giguere08, Becker13, Inoue14, ding2024} in the range $z=3$--5. We recall that the best-fit in \cite{Faucher-Giguere08} was used in our selection method in order to correct the EW$_\text{\lya}$ measurement, thus Fig.~\ref{fig:IGM_lya_forest_T} shows that our selected sample retrieves the same result.

\section{Discussion}\label{sec:discussion}

In this section, we compare our estimated redshift-dependent \lya and UV LFs with past measurements in the literature, and discuss the evolution of the QSO number densities as a function of cosmic time (Sects.~\ref{sec:Lya_LF_discussion}, \ref{sec:UV_LF_discussion}). We comment on the efficiency of our selection method to retrieve a complete and pure sample of \lya-selected QSOs (Sect.~\ref{sec:method_efficiency}), and cross-match our sample with spectroscopic surveys (Sect.~\ref{sec:Xmatch_spec}).

\subsection{\lya LF: Evolution and comparison with literature}\label{sec:Lya_LF_discussion}

As described in Sect.~\ref{sec:lya_sch_fit}, we used  a Schechter function to fit our \lya LFs (e.g., \citealt{Spinoso20}, \citetalias{Torralba-Torregrosa23}). Contrary to previous works, our data probing the faint end ($\Llya<L^*$) enables to sample three Schechter parameters with good significance. Nevertheless, the correlations between these parameters are strong, thus we choose to fix the turnover luminosity parameter $L^*$, in order to have a better understanding of the evolution of $\alpha$ and $\Phi^*$. As shown in Fig.~\ref{fig:schechter_parameters}, the normalization $\Phi^*$ decreases with redshift --- this being more evident for a fixed $L^*$. This result particularly aligns with the evolution of the total SMBH mass function predicted by semi-analytical models \citep[e.g.,][]{Fanidakis12, izquierdo-villalba2020}. In addition, in our results the faint-end slope $\alpha$ evolves toward steeper values with redshift. This trend suggests that, at lower redshift, faint AGNs tend to be proportionally less abundant than the brighter analogs ($\Llya>L^*$). Interestingly, our results about the evolution of $\Phi^*$ and $\alpha$ appear to be robust against the choice of fixing $L^*$, suggesting that the trends are likely not driven by a shift in the overall luminosity of the AGN population, but instead to be due an evolving balance between faint and bright AGN.

In general, our results are consistent with previous estimations of the AGN \lya LF in the literature. In particular, the results of this work are in agreement with \citetalias{Torralba-Torregrosa23} within the large intervals of confidence, demonstrating the consistency of our method for different datasets. The sky area in \citetalias{Torralba-Torregrosa23} is considerably smaller (1.14 deg$^2$) than the combined $\sim35$ deg$^2$ of the PAUS wide fields, and the depth of both surveys is similar. For this reason, in this work we are able to measure the LF in a range of higher \lya luminosities; pushing toward the bright end and higher redshifts. The main limiting factor at the LF bright-end is the rapid drop of bright AGNs at increasing redshifts \citep[e.g.,][]{Kulkarni19, Shen20}. The discrepancies between this work and some redshift bins in \citetalias{Torralba-Torregrosa23} can be explained by the cosmic variance due to the small area, and higher contamination in the \citetalias{Torralba-Torregrosa23} sample product of a less refined selection method.

Our estimate of the binned LF differs from that of \cite{Sobral18} and \cite{Liu22b} in the intermediate luminosity regime ($\logten(L_{{Ly}\alpha})\sim43.5$--44 erg\,s$^{-1}$). \cite{Sobral18} uses a sample selected in deep medium and narrow bands (magnitude $\sim 23.5$ and $\sim 25.5$--26 at $5\sigma$, respectively) within a smaller area of $\sim2$ deg$^{2}$. Their sample contains mainly SFGs, with a small fraction of sources that present radio or X-rays counterparts ($\sim 1.2$--$4.4\%$), suggestive of AGN activity. Our best-fits for faint end slope (ranging from $\alpha=-1.45$ to $-1.91$; mean $-1.59$) are in line with the slopes yielded by the results in \cite{Sobral18} which employed a combination of a Schechter ($\alpha=-1.7^{+0.3}_{-0.2}$) and a power-law ($A-1=-1.75\pm 0.17$), where the latter only fitted the \lya LF of the objects showing AGN radio or X-ray signatures. These fits were obtained from a sample that mainly contains objects in the \lya luminosity range of $42.5\lesssim\logten(\Llya / {\rm erg\,s}^{-1})\lesssim 43.5$ (for comparison, our luminosity range starts at $\sim 43.5$ erg\,s$^{-1}$).

In \cite{Liu22b} the \lya LF is estimated from a sample of objects selected through blind spectroscopy in HETDEX \citep{Gebhardt21}. Our lya LF at $z\sim3.24$ is significantly lower than the one of \cite{Liu22b}. Their sample contains fainter (up to $r\sim 26$) objects with typically narrower lines, that might contain significant contribution from star-formation activity and Type-II AGN \citep[e.g.,][]{Hainline11}. However, the selection method employed by \cite{Liu22b} relies on the detection of two emission lines inside a limited spectral range, which limits the completeness in  specific intervals of redshift where two or more QSO lines are not visible. In addition, their selection method is less efficient in detecting very broad lines, and thus their sample suffers from incompleteness in the brightest \lya luminosity bins. Our results capture better the exponential Schechter decay of the \lya LF at the bright end ($\logten(\Llya / {\rm erg\,s}^{-1})> 44.5$).

Finally, our bright-end of \lya LF strongly disagrees with that of \cite{Spinoso20} in the $z=3.24$ bin (their highest probed redshift). In \cite{Spinoso20} the LF is estimated from a sample of NB-selected QSOs in the $\sim 1000$ deg$^2$ of the J-PLUS survey \citep{Cenarro19} data-release one. The discrepancies with our estimation of the \lya LF at $z=3.24$ can be explained by the fact that, as stated in  \cite{Spinoso20}, their $z\sim3.24$ QSO sample is highly affected by mis-classification of  \ion{C}{IV} or \ion{C}{III} QSO lines (at $z<3$) as \lya at $z\sim3.24$. In addition, the J-PLUS data used in their work had a limiting depth of $r\lesssim21.5$, which made difficult a reliable identification of $z\sim3.24$ objects. Finally, as reported in their work, their purity computation was based on a cross-match with SDSS QSO catalogs, which become increasingly incomplete at $z>3$, likely leading to an overestimated purity for the highest-redshift bin they analyzed.

\subsection{UV LF: Evolution and comparison with literature}\label{sec:UV_LF_discussion}

The UV LF of quasars has been intensively studied in the literature, mostly focusing on spectroscopically selected samples \citep[e.g.,][]{Matsuoka18, Akiyama18, Kulkarni19, Schindler19, Niida20, Pan22}. The UV LF clearly evolves across cosmic time, and several models have been proposed to explain this evolution: from ``pure luminosity evolution'' (PLE) models, where the change in the LF is mainly driven by a shift in the luminosity distribution; to ``luminosity evolution and density evolution'' models where both the luminosity distribution and global number density of AGNs evolve \citep[see e.g.,][]{Boyle00, Ross13}.

Our best-fitting DPL parameters reflect an evolution of the UV LF (see Fig.~\ref{fig:DPL_parameters}). The normalization parameter $\Phi^*$ in general decreases with redshift, in parallel with what is found for our \lya LF Schechter fits (Sect.~\ref{sec:Lya_LF_discussion}). This decrease is in line with the evolution found by works in the literature that studied the UV LF at different redshifts, as is also made evident in Fig.~\ref{fig:eion_1450}. This trend also becomes more evident when fixing the faint-end slope. This result is consistent with a global picture in which cosmic AGN activity decreases from $z\sim2$ toward $z\sim5$ \citep[e.g.,][]{Shen20}. When left free, the faint-end slope $\beta$ shows a correlation with redshift, evolving toward flatter slopes. However, we note that the faint-end of the UV LF might not be well constrained by our data. In general, this parameter does not show consistent evolution in the literature (see Fig.~\ref{fig:DPL_parameters}). The other two DPL parameters ($\gamma, M^*$) do not present an evident evolution. In particular the bright-end slope $\gamma$ shows rather shallow values, as compared to the ones obtained by previous works. The excess in our UV LF is not significant given the uncertainties, and can be explained by cosmic variance. This excess cannot be explained by magnification bias, as this effect is predicted to become relevant at $\sim 8$~\citep{Wyithe11}. In general all the DPL parameters present significant correlations and large variance.

In \cite{Akiyama18, Niida20} and \cite{Matsuoka18}, the UV LF of quasars is computed using a combination of deep imaging data for faint AGNs ($M_{\rm UV}\gtrsim -26$) and SDSS samples of quasars for the bright-end ($M_{\rm UV}\lesssim -26$). These three works fitted significantly flatter faint-end slopes (mean $\beta=-1.25$; see Fig.~\ref{fig:DPL_parameters}, with no significant evolution with redshift. On the other hand, \cite{Kulkarni19} obtained steeper faint-end slopes, by only employing QSOs from SDSS and BOSS, which in general are more luminous than the sample of HSC deep imaging. In order to try to reconcile our results with the faint-end slope constrained by HSC, we fit a DPL model to our UV LFs using a fixed $\beta=-1.25$. The fitted curve is displayed in Fig.~\ref{fig:multi_UV_LF} as a teal solid line, and it is generally compatible with our binned LF, showing that our data does not tightly constrain the faint-end slope. The fits yield slightly fainter values for the turnover magnitude $M^*$, which are more in agreement with the fainter values for $M^*$ derived from the HSC+SDSS data. By fixing $\beta=-1.25$ we also obtain better constrained values of the bright-end slope, and more in line with the HSC+SDSS determinations. Moreover, the anticorrelation of $\Phi^*$ with redshift increases, reflecting an evolution consistent with that of the Ly$\alpha$ LF (see Sect.~\ref{sec:lya_sch_fit}). The fact that fixing $\beta=-1.25$ yields fainter turnover magnitudes might mean that our UV LF only covers a specific $M_{\rm UV}$ range (i.e., $M_{\rm UV}\sim M^*$), hence hindering our possibility to well constrain the faint-end of the power-law LF, which is instead well-sampled in the HSC data.

\subsection{Efficiency of the selection method}\label{sec:method_efficiency}

The selection method employed in this work (see Sect.~\ref{sec:selection_method}) effectively selects quasars based on strong line emission, selected mainly through NB flux excess. We do not apply any direct color cut in our selection, which removes biases that affect most spectroscopic samples that rely on photometric pre-selection on color-color diagrams \citep[e.g.,][]{Schneider10, Ross13, Lyke20}. Selection methods based on QSO variability have been proved to be capable to obtain more pure and complete samples than color-based selections, however loose color cuts are often also applied in order to minimize contamination from stars \citep[e.g.,][]{Palanque-Delabrouille11}. The problems introduced by color cuts especially affect the redshift range $2.5\lesssim z \lesssim 3.5$, where most QSOs occupy the sample region as the stellar locus in optical color-color planes \citep{Fan99}, and the selection functions become inefficient \citep[see e.g.,][]{Richards06, Ross13}. Moreover, variability in quasars anti-correlates with continuum luminosity and redshift, and the variability of emission lines is found to be only $\sim 10\%$ of the variability of the continuum \citep{Meusinger11}. In \cite{Schindler17} it is shown that the classical QSO selection methods employed by SDSS are prone to miss a significant part of $z>3$ quasars in the brightest end, mainly due to the inefficiency of the color-based selection.

\begin{figure}
    \centering
    \includegraphics[width=\linewidth]{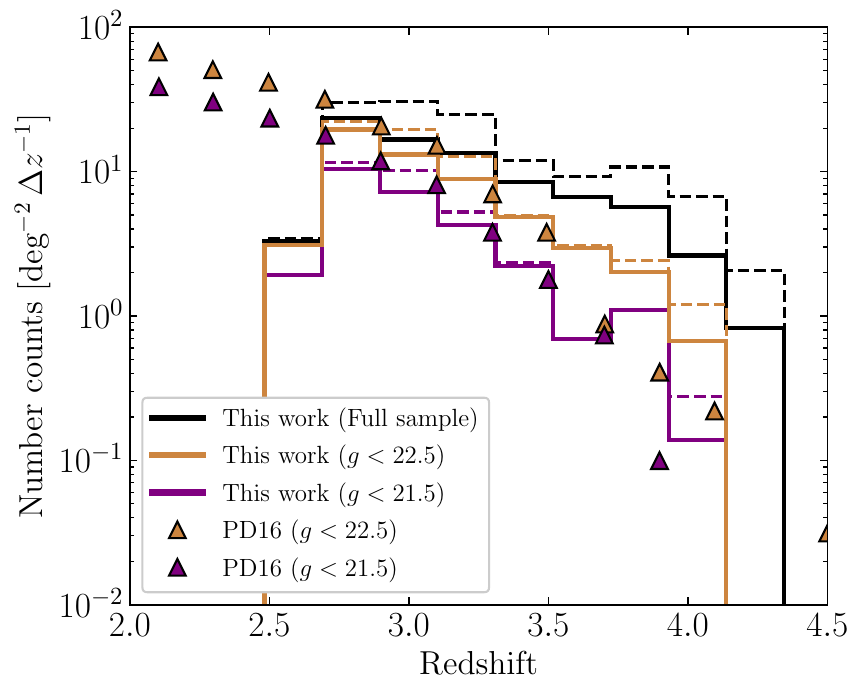}
    \caption{Differential number counts of QSOs per unit area of sky. The dashed histograms show our complete selected sample, the solid histograms show the distribution of our visually selected ``golden sample''. We compare with the spectroscopic sample of \protect\cite{Palanque-Delabrouille16} (PD16), selected through QSO variability. Our number counts are significantly higher than those of PD16, especially at the highest bins of redshift.}
    \label{fig:g_LF}
\end{figure}

In Fig.~\ref{fig:g_LF} we show the number counts per unit area in different $g$-band cuts for $2.75<z<4.3$. Our number counts are in line with those of \cite{Palanque-Delabrouille16} for $z<3.5$, from a variability-selected QSO sample. For higher redshifts ($z>3.5$), our method appears to be more efficient selecting fainter QSOs ($g>22.5$), due to the intrinsic incompleteness of the SDSS data at such magnitudes \citep[the completeness of SDSS imaging data rapidly decays for $r>22$;][]{Aihara11}.

\subsection{Cross-match with spectroscopic surveys}\label{sec:Xmatch_spec}

\begin{figure}
    \centering
    \includegraphics[width=\linewidth]{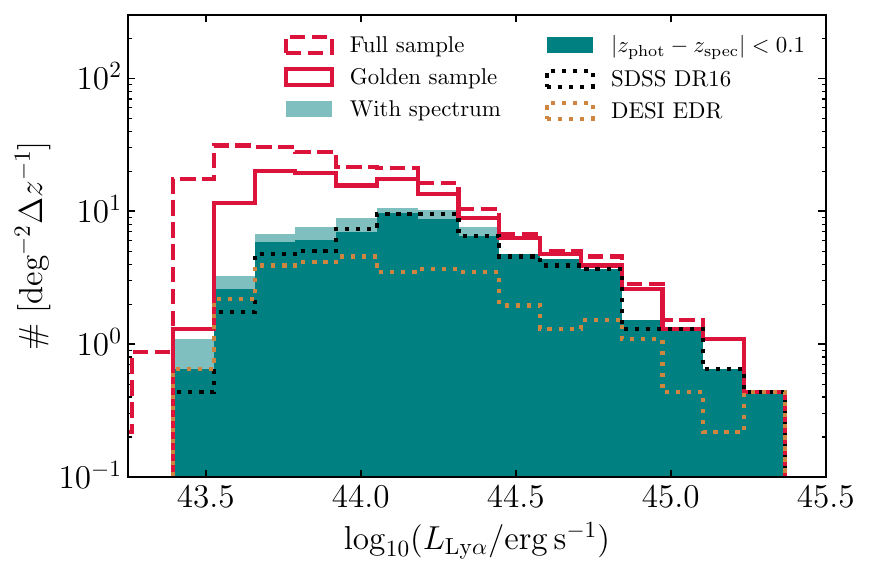}
    \includegraphics[width=\linewidth]{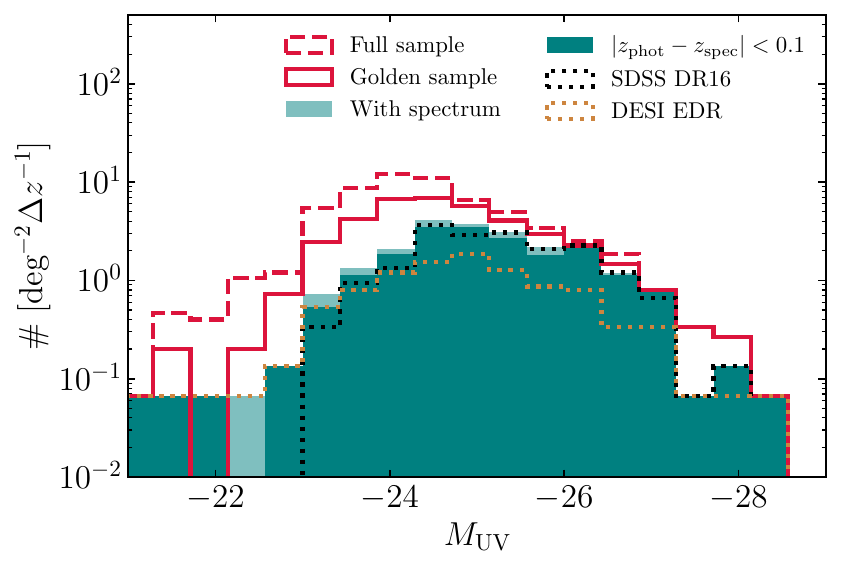}
    \caption{Differential number counts per square degree as a function of \lya luminosity (top) and $M_{\rm UV}$ (bottom). We show the counts for the full sample (dashed red line), the visual "golden sample" (solid red line) and the sub-sample with confirmed spectroscopic redshift in either SDSS, HETDEX or DESI (teal solid histograms). Most of the objects in the spectroscopic sub-sample have matching photometric redshifts ($\sim 90\%$).}
    \label{fig:spec_xmatch_histogram}
\end{figure}

We cross-match our catalog of selected LAE candidates with the catalogs of SDSS DR18 \citep{Almeida23}, DESI EDR \citep{DESI23} and HETDEX Public Source Catalog 1 \citep{Mentuch-Cooper23}. The match with these catalogs yield a sub-sample of 370 objects with spectroscopic counterpart: 229, 132 and 9 in SDSS, DESI and HETDEX, respectively. The comparison between our measured redshifts and the spectroscopic redshifts measured by the respective surveys is shown in Fig.~\ref{fig:redshift_phot_spec}, and discussed in Sect.~\ref{sec:redshift_correction} above.

In Fig.~\ref{fig:spec_xmatch_histogram} we show the differential number counts per unit area in our selected sample as compared to the sub-sample having spectroscopic counterparts. While the overlap between the coverage of SDSS DR16 and all three PAUS wide fields is total, there is only partial overlap between those and HETDEX Public Catalog 1 and DESI EDR. Our sample is only slightly more complete than that of SDSS DR16 for $\logten(\Llya / {\rm erg\,s}^{-1})\gtrsim 44.25$, and contains a significantly larger number of objects for fainter \lya luminosities ($\sim 1$ dex larger at $\logten(\Llya / {\rm erg\,s}^{-1})\approx 43.75$). On the other hand, the ratio between the number density of our sample and the QSO sample in the DESI EDR appears to be constant down to $\logten(\Llya / {\rm erg\,s}^{-1})\approx 43.5$.

%%%%%%%%%%%%%%%%%%%%%%%%%%%%%%%%%%%%%%%%%%%%%%%%%%%%%%%%%%%%%%%%%%%%
\section{Summary}\label{sec:summary}

In this work we have presented a method to select QSOs through \lya emission lines in the PAU Survey filter system. This method is a refinement of that developed in \cite{Torralba-Torregrosa23}.
We built mock catalogs of the target sample, QSOs with strong \lya emission, and the expected contaminants: that is QSOs at redshifts where \lya cannot be observed with our NB filters, and low-$z$ galaxies.

We probed a total effective area of $35.68$ deg$^2$ in three PAUS wide fields (PAUS W1, W2 and W3), and obtained a sample of 915 \lya selected QSOs, in the redshift range $2.7<z<5.3$, with \lya luminosities of $43.5 < \logten (\Llya / {\rm erg\,s}^{-1}) < 45.5$. Our method obtains, in the first place, a sample of candidates selected through prominent flux excess in one or more NBs, that can be compatible with QSO lines at such redshifts. Then we used a neural-network classifier to obtain a purer sample, trained with mock catalogs. This classifier is able to correctly pick up \lya emitting QSOs from our pre-selected sample, with an accuracy of $\approx 95\%$, while maintaining a low false-relative rate ($\approx 7\%$).

A random forest regressor was used, trained with the same mock catalogs in order to obtain precise redshifts. The determination of the redshift using the \lya detection NB only is slightly inaccurate, since the intrinsic width of the observed wavelength coverage of the NBs correspond to a redshfit uncertainty of $\Delta z \approx 0.12$. Moreover, since QSO lines can be fairly wide in velocity space (up to $\sim 5000$ km\,s$^{-1}$), the selected NB could not correspond to the peak of the \lya line, in addition to the expected bias of the QSO \lya redshift with respect to the systemic redshift. Our algorithm effectively obtains more precise redshifts, when compared to the spectroscopically measured redshifts.

The purity of our sample reaches $\approx 90\%$ at $\logten (\Llya / {\rm erg\,s}^{-1})\approx 44.5$ for the lowest redshift intervals ($2.75<z<3.2$) and at $\logten (\Llya / {\rm erg\,s}^{-1})\approx 44$ for $3.2<z<4.2$, as estimated using our mocks. The purity of the sub-sample with spectroscopic counterparts (cross-matching with SDSS, DESI and HETDEX) is $\sim 88\%$ for the full selected sample, and $\approx 91\%$ ($\approx 100\%$) for \lya luminosity $\logten (\Llya / {\rm erg\,s}^{-1}) > 44$ ($\logten (\Llya / {\rm erg\,s}^{-1}) > 44.5$).
The main source of contamination, as predicted from our mocks and confirmed by the spectroscopic cross-matched sub-sample, are lower-$z$ QSOs ($z<2.7$). In particular, the strong \ion{C}{IV} QSO line is the main source of misidentification with \lya.

We estimated the \lya and UV ($M_{\rm UV}$, at rest-frame 145 nm) LFs. For this task, we obtained estimations of purity and number count corrections over a grid of $r_{\rm synth}$ (a synthetic broad-band from the combination of NBs, mimicking the classical band $r$) and $\Llya$, in different intervals of redshift. Both the \lya and UV LFs show evident evolution, with declining normalization parameter $\Phi^*$, fitting a Schechter function, and a double power-law, respectively. The faint-end of the \lya LF shows a clear correlation with redshift in the probed interval, suggesting that the number density of bright quasars declines faster than the faint counterparts, in agreement with the SMBH growth models at $z\gtrsim 2$.

We obtained composite QSO spectra from NB data, for different intervals of redshift. We measured a UV continuum slope of $\beta_{\rm UV}=-1.55\pm 0.02$. We employ the composite spectra to measure the mean IGM transmision of the \lya forest, obtaining results compatible with previous spectroscopic determinations.

The total \lya luminosity density, obtained from the integral of the \lya LF, shows a rapid decline with increasing redshift, being sub-dominant by several orders of magnitude already by $z\approx 4$, with respect to that of star-formation \lya emission.

\section*{Data availability}

The catalog of selected objects elaborated in this work will be available at \href{https://cosmohub.pic.es/home}{CosmoHub} (\citealt{CosmoHub1, CosmoHub2}), together with the whole PAUS public data release, through this link: \href{https://cosmohub.pic.es/catalogs/319}{https://cosmohub.pic.es/catalogs/319}. More information on the data release, as well as access to the PAUS database and raw and reduced images, can be found at \href{https://pausurvey.org/public-data-release/}{https://pausurvey.org/public-data-release/}. The stacking code \texttt{stonp} is publicly available at \href{https://github.com/PAU-survey/stonp}{https://github.com/PAU-survey/stonp}.

%%%%%%%%%%%% ACKNOWLEDGEMENTS %%%%%%%%%%%%%%%
\begin{acknowledgements}

The authors acknowledge the insightful feedback from the scientific referee, which has significantly contributed to enhancing the quality of this paper.

This work has been funded by project PID2019-109592GBI00/AEI/10.13039/501100011033 from the Spanish Ministerio de Ciencia e Innovación (MCIN)—Agencia Estatal de Investigación, by the Project of Excellence Prometeo/2020/085 from the Conselleria d’Innovació Universitats, Ciència i Societat Digital de la Generalitat Valenciana.

This work is part of the research Project PID2023-149420NB-I00 funded by MICIU/AEI/10.13039/501100011033 and by ERDF/EU.

This work is also supported by the project of excellence PROMETEO CIPROM/2023/21 of the Conselleria de Educación, Universidades y Empleo (Generalitat Valenciana).

The authors acknowledge the financial support from the MCIN with funding from the European Union NextGenerationEU and Generalitat Valenciana in the call Programa de Planes Complementarios de I+D+i (PRTR 2022). Project (VAL-JPAS), reference ASFAE/2022/025.

PR and DS acknowledge the support by the Tsinghua Shui Mu Scholarship. PR, DS and ZC acknowledge the funding of the National Key R\&D Program of China (grant no. 2023YFA1605600), the science research grants from the China Manned Space Project with No. CMS-CSST2021-A05, and the Tsinghua University Initiative Scientific Research Program (No. 20223080023). PR acknowledges additional funding from the National Science Foundation of China (grant no. 12350410365).

EG acknowledges grants from Spain Plan Nacional (PGC2018-102021-B-100) and Maria de Maeztu (CEX2020-001058-M).

The PAU Survey is partially supported by MINECO under grants CSD2007-00060, AYA2015-71825, ESP2017-89838, PGC2018-094773, PGC2018-102021, PID2019-111317GB, SEV-2016-0588, SEV-2016-0597, MDM-2015-0509 and Juan de la Cierva fellowship and LACEGAL and EWC Marie Sklodowska-Curie grant No 734374 and no.776247 with ERDF funds from the EU Horizon 2020 Programme, some of which include ERDF funds from the European Union. IEEC and IFAE are partially funded by the CERCA and Beatriu de Pinos program of the Generalitat de Catalunya. Funding for PAUS has also been provided by Durham University (via the ERC StG DEGAS-259586), ETH Zurich, Leiden University (via ERC StG ADULT-279396 and Netherlands Organisation for Scientific Research (NWO) Vici grant 639.043.512), University College London and from the European Union's Horizon 2020 research and innovation programme under the grant agreement No 776247 EWC. The PAU data center is hosted by the Port d'Informaci\'o Cient\'ifica (PIC), maintained through a collaboration of CIEMAT and IFAE, with additional support from Universitat Aut\`onoma de Barcelona and ERDF. We acknowledge the PIC services department team for their support and fruitful discussions.

EG acknowledges grants from Spain Plan Nacional (PGC2018-102021-B-100) and Maria de Maeztu (CEX2020-001058-M). JC acknowledges support from the grant PID2021-123012NA-C44 funded by MCIN/AEI/ 10.13039/501100011033 and by “ERDF A way of making Europe”. FJC is supported by grants PID2022-141079NB and CEX2020-001058-M funded by MCIN/AEI/10.13039/501100011033. ME acknowledges funding by MCIN with funding from European Union NextGenerationEU (PRTR-C17.I1) and by Generalitat de Catalunya. IFT participation is supported by the grant PID2021-123012NB-C43P funded by MCIN/AEI /10.13039/501100011033. H. Hildebrandt is supported by a DFG Heisenberg grant (Hi 1495/5-1), the DFG Collaborative Research Center SFB1491, as well as an ERC Consolidator Grant (No. 770935). IFAE participation is supported by grant PID2021-123012NB funded by MCIN/AEI/10.13039/501100011033.  CIEMAT participation is supported by the grant PID2021-123012NB-C42P funded by MCIN/AEI /10.13039/501100011033.

\end{acknowledgements}

%%%%%%%%%%%%%%%%%%%% REFERENCES %%%%%%%%%%%%%%%%%%
\bibliographystyle{aa}
\bibliography{my_bibliography}

%%%%%%%%%%%%%%%%%%%%%%%%%%%%%%%%%%%%%%%%%%%%%%%%%%%%%%%%%%%%%%%%%%%%
%%%%%% APPENDICES %%%%%%%
\appendix
%%%%%%%%%%%%%%%%%%%%%%%%%%%%%%%%%%%%%%%%%%%%%%%%%%%%%%%%%%%%%%%%%%%%

\section{Ly$\alpha$ redshift of the PAUS narrow bands}

In Table~\ref{tab:NB_lya_redshifts} we display the list of NBs used to select \lya emitting candidates, the associated \lya redshift of each NB and the number of candidates selected.

\begin{table}
\centering
    \caption{NB filters used to select \lya emitting quasars.}
    \label{tab:NB_lya_redshifts}
\begin{tabular}{ccccc}
\toprule
NB ID & Name$^a$ & 5$\sigma$ magAB$^b$ & \lya $z$ ($\pm0.06$)$^c$ & \# Selected$^d$\\
\midrule
 1 & NB455 & 22.18 & 2.76 & 103\\
 2 & NB465 & 22.17 & 2.84 & 120\\
 3 & NB475 & 22.27 & 2.91 & 88\\
 4 & NB485 & 22.41 & 2.99 & 117\\
 5 & NB495 & 22.36 & 3.08 & 80\\
 6 & NB505 & 22.30 & 3.16 & 66\\
 7 & NB515 & 22.24 & 3.25 & 65\\
 8 & NB525 & 22.30 & 3.32 & 29\\
 9 & NB535 & 22.27 & 3.41 & 32\\
 10 & NB545 & 22.31 & 3.50 & 13\\
 11 & NB555 & 22.11 & 3.58 & 26\\
 12 & NB565 & 22.26 & 3.66 & 23\\
 13 & NB575 & 22.31 & 3.73 & 20\\
  14 & NB585 & 22.13 & 3.82 & 25\\
  15 & NB595 & 22.06 & 3.90 & 17\\
  16 & NB605 & 22.25 & 3.98 & 4\\
  17 & NB615 & 22.34 & 4.06 & 8\\
  18 & NB625 & 22.21 & 4.15 & 10\\
  19 & NB635 & 22.23 & 4.23 & 4\\
  20 & NB645 & 22.39 & 4.31 & 2\\
  21 & NB655 & 22.38 & 4.39 & 1\\
  22 & NB665 & 22.38 & 4.47 & 6\\
  23 & NB675 & 22.35 & 4.56 & 0\\
  24 & NB685 & 22.14 & 4.64 & 1\\
  25 & NB695 & 22.12 & 4.72 & 2\\
  26 & NB705 & 22.27 & 4.81 & 1\\
  27 & NB715 & 22.37 & 4.89 & 2\\
  28 & NB725 & 22.04 & 4.97 & 0\\
  29 & NB735 & 21.95 & 5.05 & 2\\
  30 & NB745 & 22.07 & 5.13 & 0\\
  31 & NB755 & 21.93 & 5.21 & 1\\
  % NB765 & 21.80 & 5.31 & \\
  % NB775 & 21.8493982255349 & 5.370972385598065\\
  % NB785 & 21.751670551922995 & 5.451586368011055\\
  % NB795 & 21.77419097590486 & 5.538781083682249\\
  % NB805 & 21.90328552953494 & 5.618572474437964\\
  % NB815 & 22.104853667316206 & 5.703299415137331\\
  % NB825 & 21.877778576283884 & 5.784735989207597\\
  % NB835 & 21.595234626085396 & 5.875221071507893\\
  % NB845 & 21.646755849604688 & 5.956657645578158\\
\bottomrule
\end{tabular}
\tablefoot{$^a$The name of the filters is composed of ``NB'' + the approximate effective wavelength in nm. $^b$Average limit magnitude at the 5$\sigma$ level. $^c$Interval of \lya redshifts probed. The amplitude of this interval is defined by the FWHM of the filters. $^d$Number of selected \lya emitting quasars in the final sample by each NB.}
\end{table}

\section{ANN classifier details}\label{sec:NN_class_details}

\begin{figure*}
    \centering
    \includegraphics[width=0.49\linewidth]{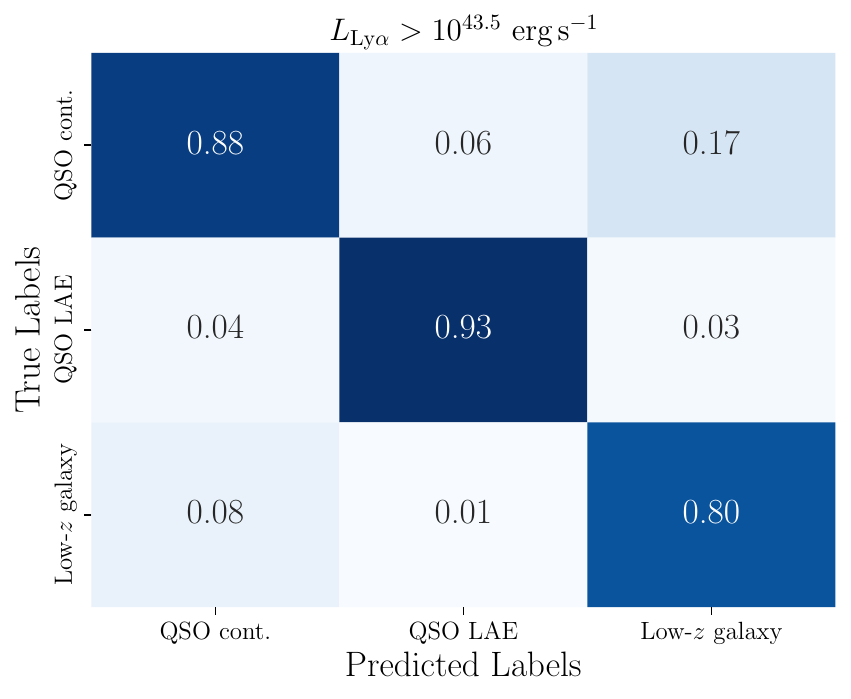}
    \includegraphics[width=0.49\linewidth]{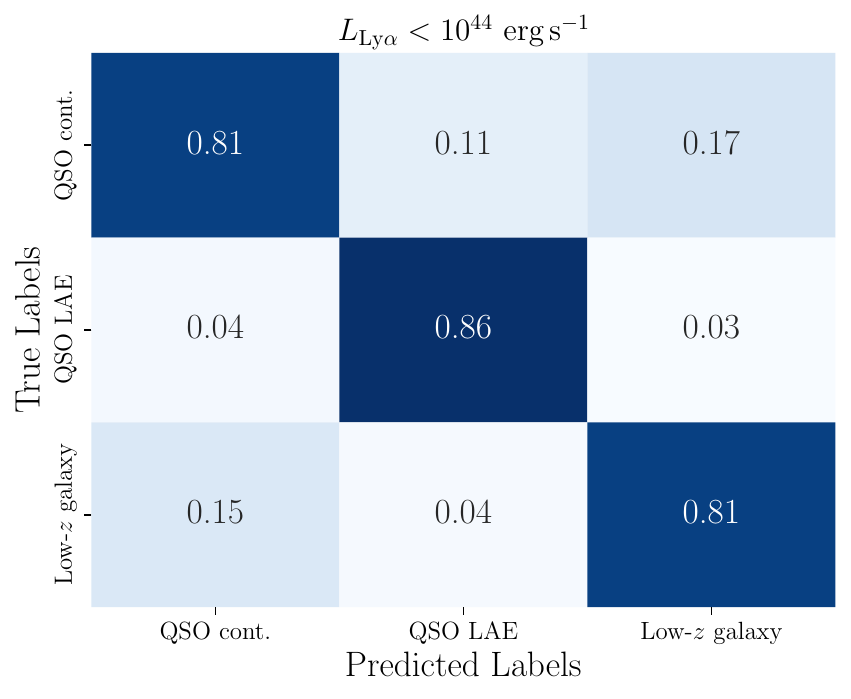}
    \includegraphics[width=0.49\linewidth]{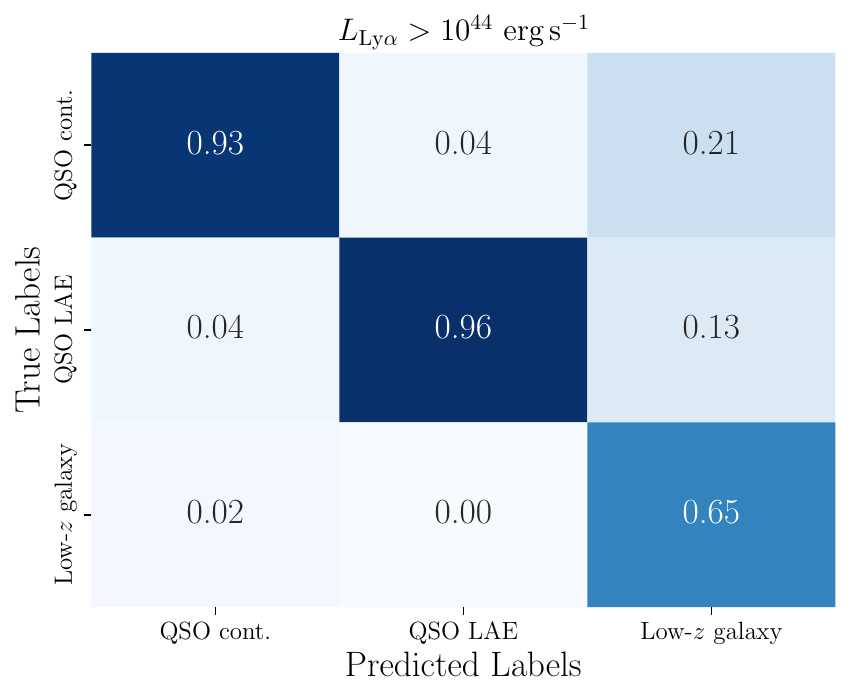}
    \caption{Confusion matrix for the ANN classifier used to select LAEs. The three classes considered are: QSOs selected by a feature other than \lya (``QSO cont.''), QSOs correctly selected by their \lya line, in the correct redshift (``QSO LAE''), and low-$z$ ($z\approx 0$--$2$) galaxy contaminants (``Low$-z$ galaxy''). The annotations in the cells are the normalized counts so the vertical columns add up to 1. We show the confusion matrices for the full test set (top, $\Llya>10^{43.5}$ erg\,s$^{-1}$) and splitting in low (middle, $\Llya<10^{44}$ erg\,s$^{-1}$) and high luminosity bins (bottom, $\Llya>10^{44}$ erg\,s$^{-1}$). This figure shows that the ANN effectively selects QSO LAEs with an accuracy of 93\% for the general sample, while maintaining a low false negative rate (6\% and 1\% in the case of objects labeled as QSO cont. and low-$z$ galaxies, respectively.)}
    \label{fig:conf_matrix_NN}
\end{figure*}

For the machine learning models used in this work, we use the python library Sci-kit learn \citep{scikit-learn}. The ANN classifier is built using the \texttt{MLPClassifier} method. We performed a grid-search of optimal parameters with \texttt{GridSeachCV}, over an extense grid of parameters. The ANN source classifier optimal parameters are:

\begin{itemize}
\item[{}] \texttt{hidden\_layer\_sizes = (60, 60, 60),\\
batch\_size = 64,\\
alpha = 1e-4,\\
learning\_rate = 'adaptive',\\
max\_iter = 1e4,\\
solver = 'adam'}.
\\
\end{itemize}

In Fig.~\ref{fig:conf_matrix_NN} we show the confusion matrices of the three classes, in diferent bins of measured \lya luminosity.

For the RF redshift regressor we use the method \texttt{RandomForestRegressor}, and perform a grid-search using the same method employed for the NN. We obtain the following optimal parameters:

\begin{itemize}
\item[{}]\texttt{min\_samples\_split = 3,\\
min\_samples\_leaf = 2,\\
max\_depth = 20,\\
bootstrap = True
}
\end{itemize}

\section{2D purity and completeness}

\begin{figure*}
    \centering
    \includegraphics[width=0.49\linewidth]{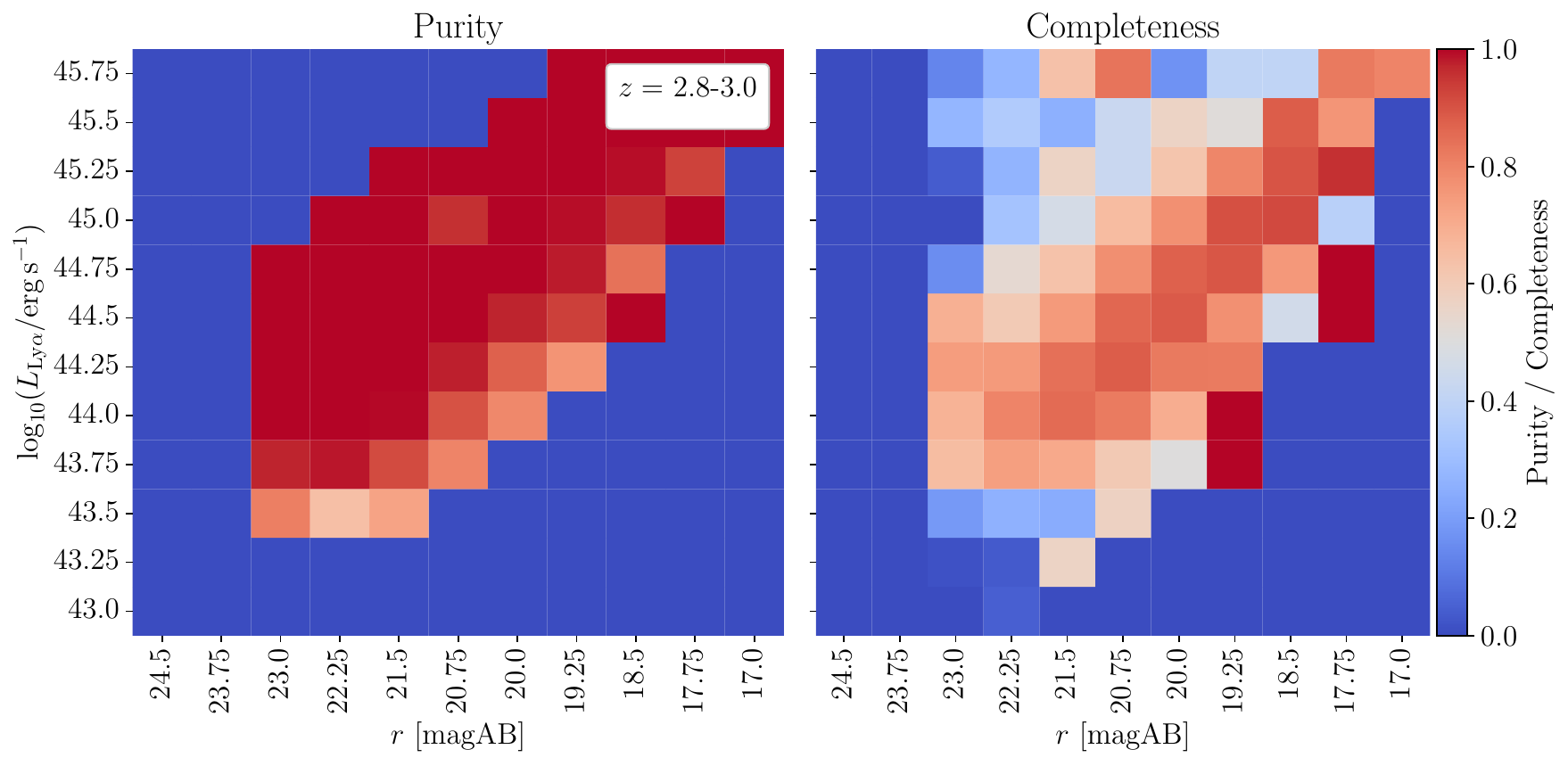}
    \includegraphics[width=0.49\linewidth]{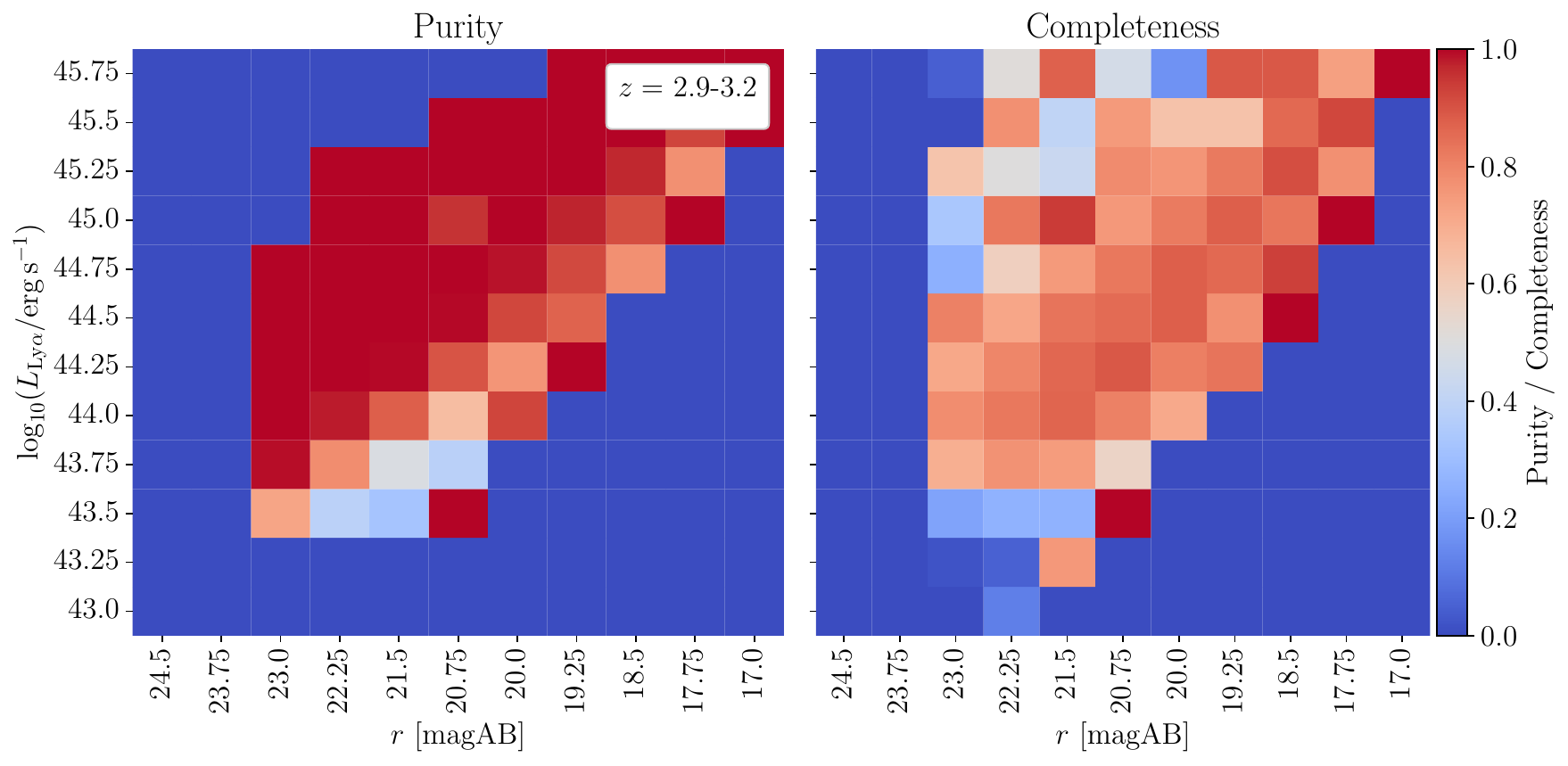}
    \includegraphics[width=0.49\linewidth]{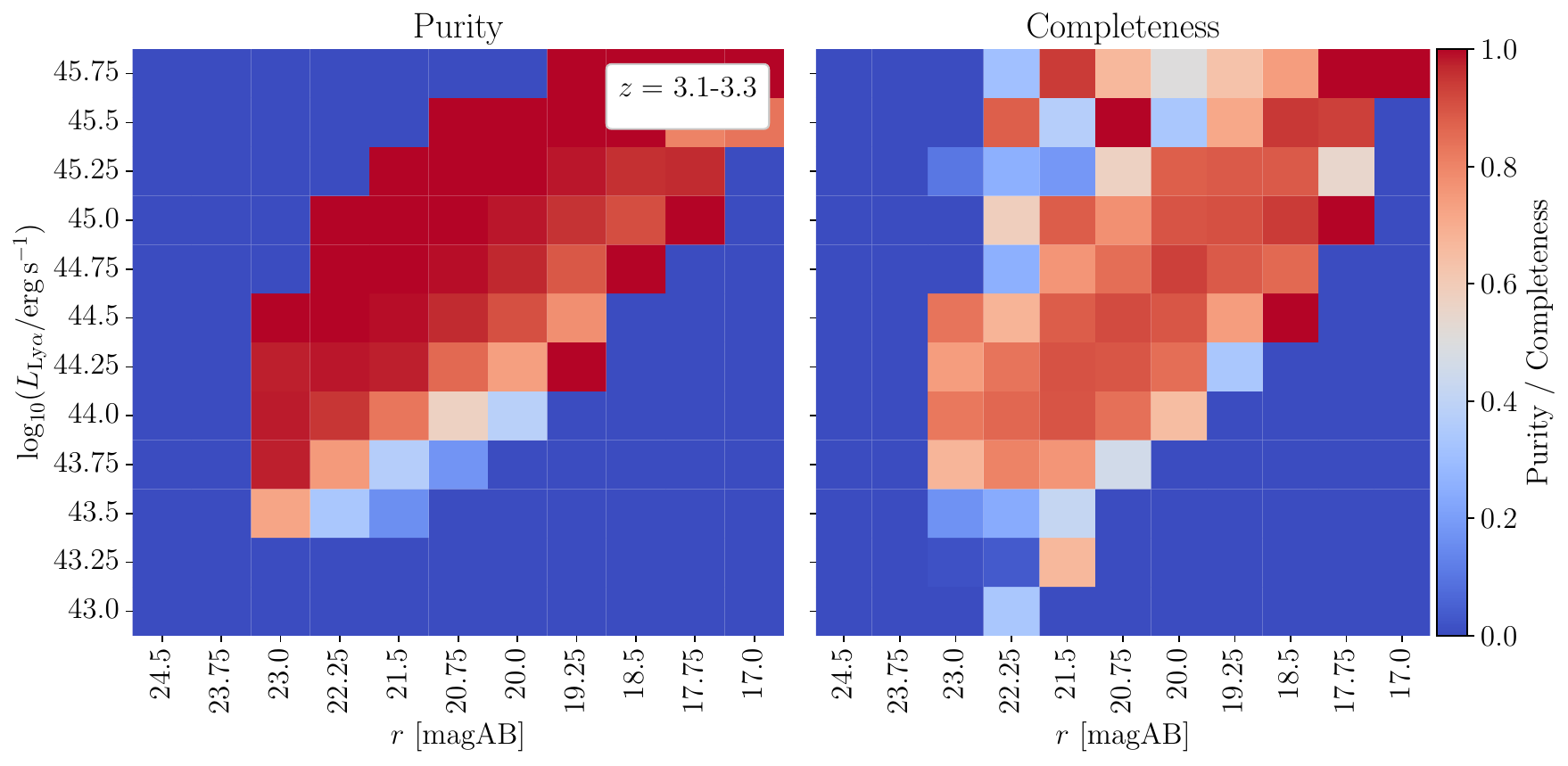}
    \includegraphics[width=0.49\linewidth]{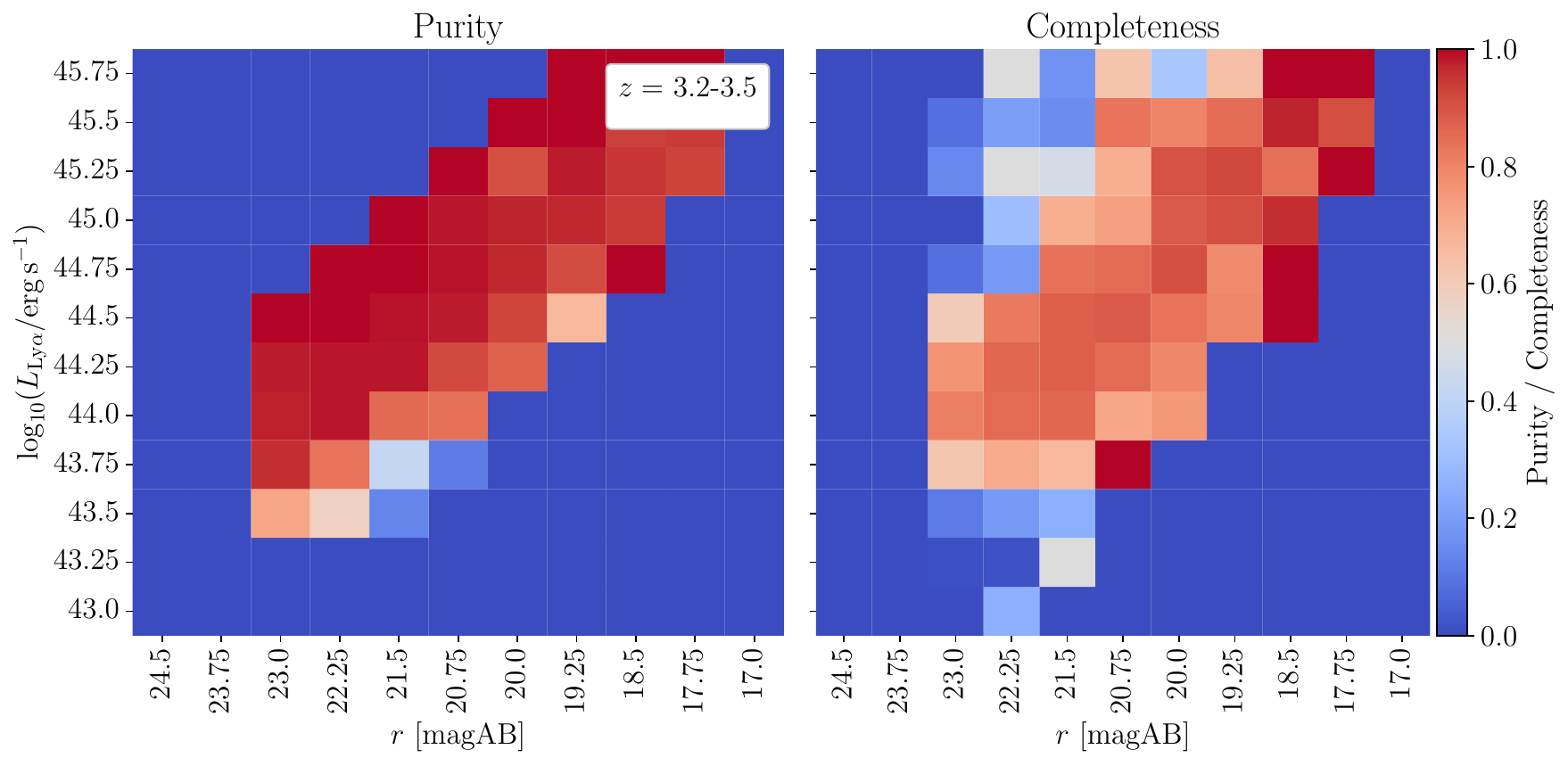}
    \includegraphics[width=0.49\linewidth]{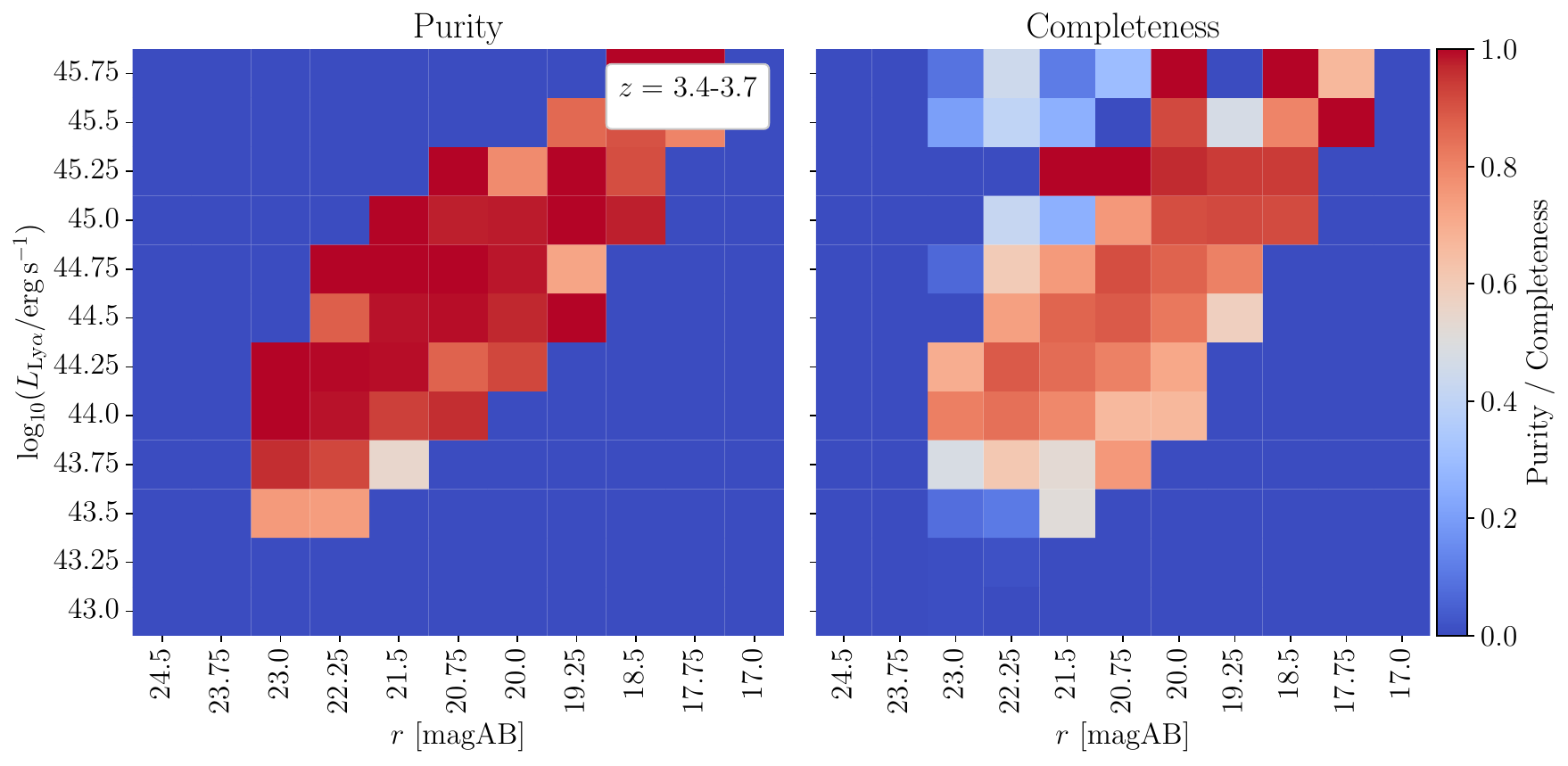}
    \includegraphics[width=0.49\linewidth]{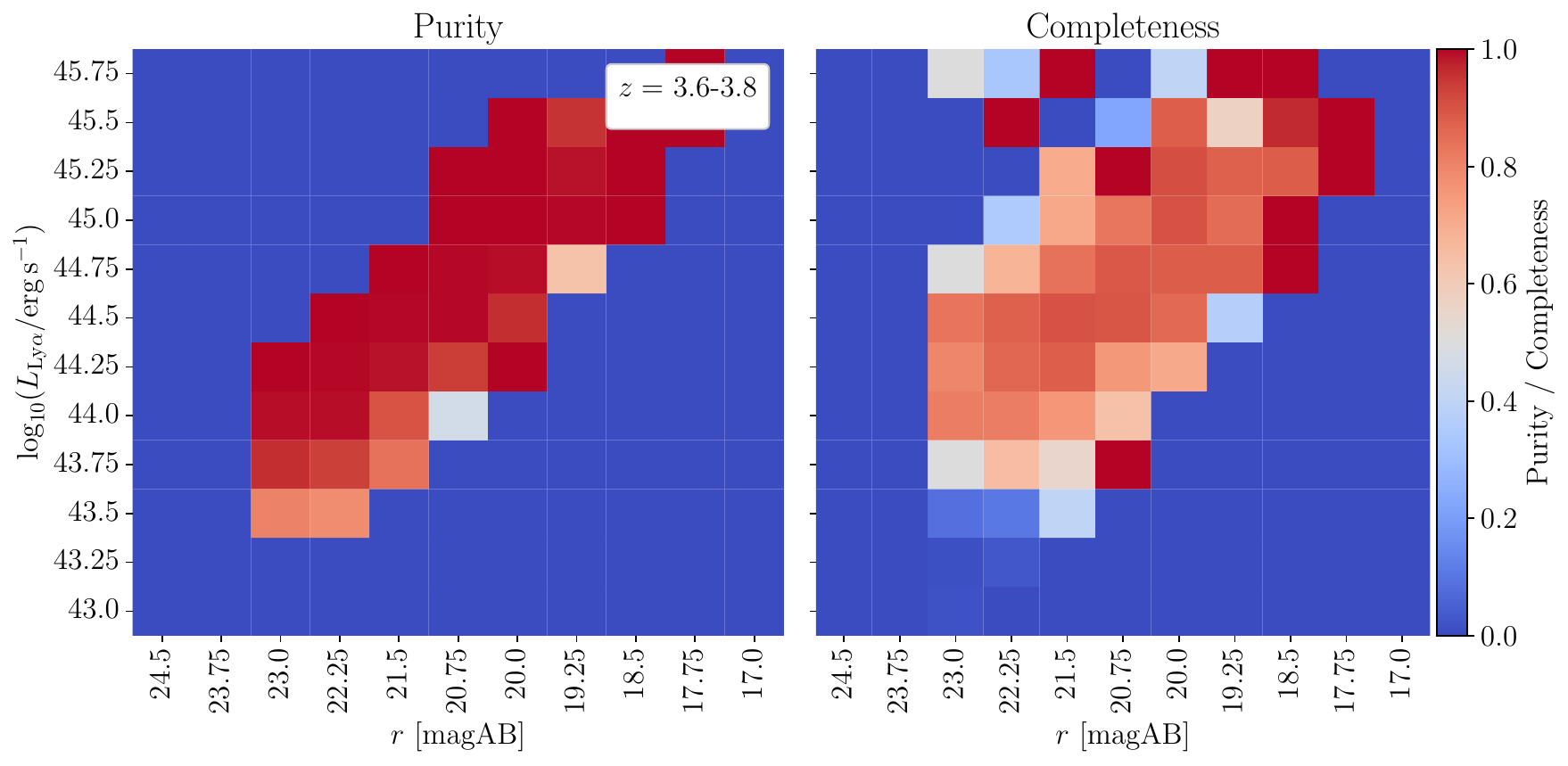}
    \includegraphics[width=0.49\linewidth]{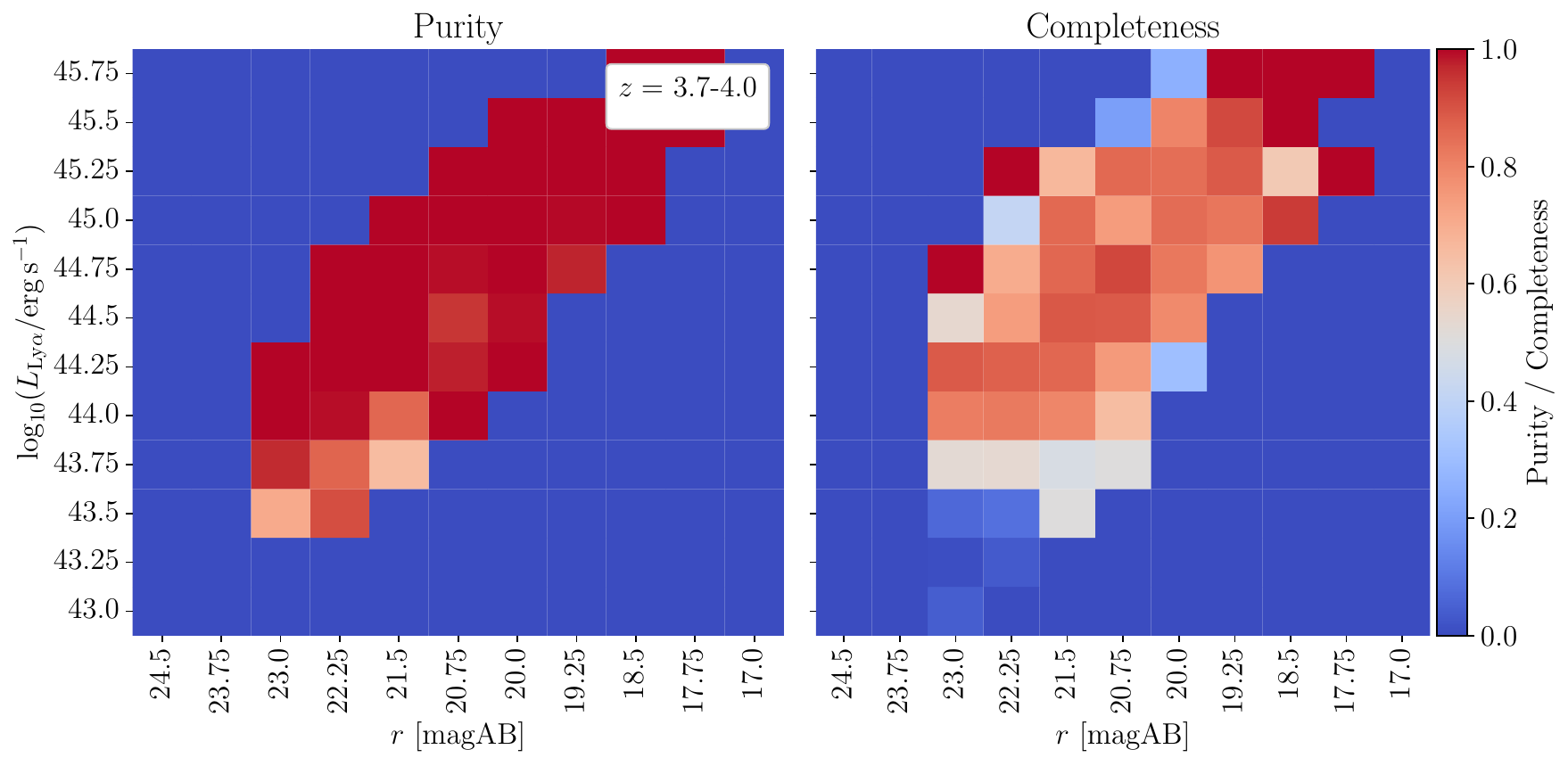}
    \includegraphics[width=0.49\linewidth]{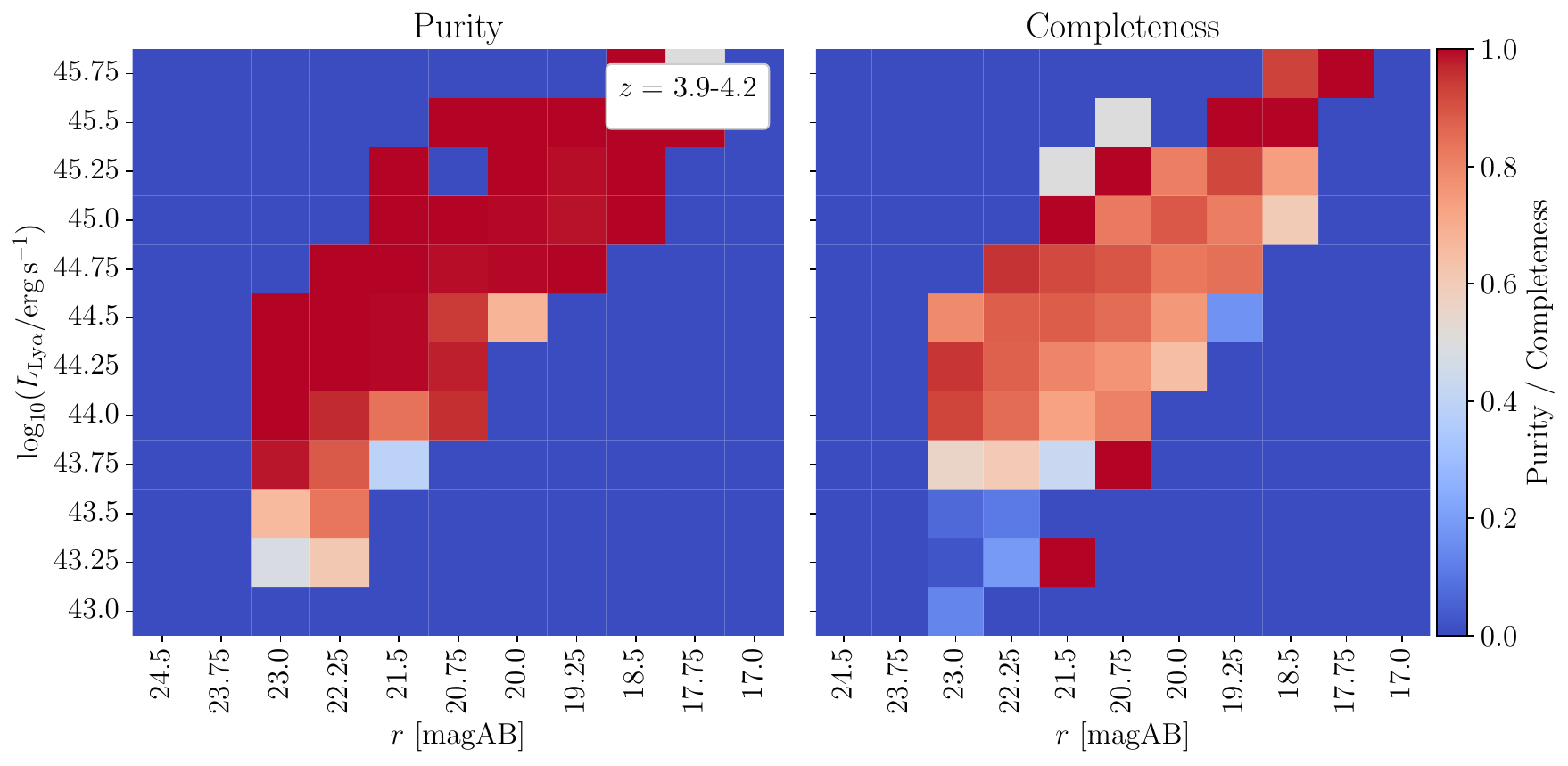}
    \caption{2D maps of purity and completeness in bins of $\Llya$ and magnitude $r$. The values displayed in Fig.~\ref{fig:1d_puricomp} are obtained by collapsing these 2D maps over the horizontal axis.}
    \label{fig:2D_purity_and_comp}
\end{figure*}

In Fig.~\ref{fig:2D_purity_and_comp} we display the purity and completeness of the selection method in bins of $r$ and measured $\Llya$. These 2D maps are obtained by applying the selection method on the mocks (Sect.~\ref{sec:mocks}) and comparing the selected sample with the parent sample using Eqs.~\ref{eq:purity}~and~\ref{eq:completeness}.

\section{Schechter and double power-law fit parameters}

In Tables~\ref{tab:sch_params} and~\ref{tab:dpl_params} we list the best-fit parameters to a Schechter function for the \lya LF, and double powerlaw for the UV LF, respectively. For the \lya LF Schechter fit we list the results of fitting with all three free Schechter parameters, and fixing $\logten (L^* / {\rm erg\,s^{-1}})=45$, as discussed in Sect.~\ref{sec:lya_sch_fit}. Similarly, for the UV LF fit, we list the results of leaving all four DPL parameters free, and fixing the faint-end slope $\beta=-1.25$, as discussed in Sect.~\ref{sec:dpl_fit}

\begin{table}
\centering
    \caption{Best-fit Schechter parameters for the \lya LF.}
    \label{tab:sch_params}
\resizebox{\linewidth}{!}{
\begin{tabular}{ccccc}
\toprule

Redshift & $\logten(\Phi^*/{\rm Mpc}^{-3})$ & $\logten (L^*/{\rm erg\,s}^{-1})$ & $\alpha$ \\

\midrule

2.76 - 2.99 & $-6.48^{+0.24}_{-0.28}$ & $44.89^{+0.21}_{-0.19}$ & $-1.44^{+0.18}_{-0.16}$ \\
2.91 - 3.16 & $-6.63^{+0.20}_{-0.28}$ & $44.98^{+0.21}_{-0.17}$ & $-1.30^{+0.17}_{-0.17}$ \\
3.08 - 3.32 & $-6.81^{+0.23}_{-0.30}$ & $45.06^{+0.22}_{-0.20}$ & $-1.34^{+0.18}_{-0.17}$ \\
3.25 - 3.50 & $-7.00^{+0.38}_{-0.42}$ & $45.05^{+0.26}_{-0.28}$ & $-1.60^{+0.26}_{-0.22}$ \\
3.41 - 3.66 & $-6.80^{+0.45}_{-0.70}$ & $44.73^{+0.44}_{-0.31}$ & $-1.63^{+0.39}_{-0.34}$ \\
3.58 - 3.82 & $-6.80^{+0.43}_{-0.57}$ & $44.77^{+0.39}_{-0.31}$ & $-1.62^{+0.32}_{-0.26}$ \\
% \midrule
% 2.76 - 2.99 & $-6.98^{+0.13}_{-0.13}$ & $45.18^{+0.15}_{-0.15}$ & -1.69 (fixed) \\
% 2.91 - 3.16 & $-7.32^{+0.13}_{-0.09}$ & $45.34^{+0.11}_{-0.17}$ & -1.69 (fixed) \\
% 3.08 - 3.32 & $-7.42^{+0.13}_{-0.10}$ & $45.34^{+0.11}_{-0.17}$ & -1.69 (fixed) \\
% 3.25 - 3.50 & $-7.24^{+0.20}_{-0.16}$ & $45.20^{+0.19}_{-0.24}$ & -1.69 (fixed) \\
% 3.41 - 3.66 & $-6.99^{+0.28}_{-0.27}$ & $44.85^{+0.31}_{-0.30}$ & -1.69 (fixed) \\
% 3.58 - 3.82 & $-6.90^{+0.18}_{-0.18}$ & $44.80^{+0.18}_{-0.18}$ & -1.69 (fixed) \\
\midrule
2.76 - 2.99 & $-6.62^{+0.08}_{-0.09}$ & 45 (fixed) & $-1.51^{+0.08}_{-0.09}$ \\
2.91 - 3.16 & $-6.65^{+0.09}_{-0.10}$ & 45 (fixed) & $-1.30^{+0.10}_{-0.11}$ \\
3.08 - 3.32 & $-6.72^{+0.10}_{-0.11}$ & 45 (fixed) & $-1.28^{+0.11}_{-0.12}$ \\
3.25 - 3.50 & $-6.90^{+0.13}_{-0.15}$ & 45 (fixed) & $-1.54^{+0.14}_{-0.16}$ \\
3.41 - 3.66 & $-7.19^{+0.18}_{-0.23}$ & 45 (fixed) & $-1.81^{+0.18}_{-0.22}$ \\
3.58 - 3.82 & $-7.13^{+0.13}_{-0.15}$ & 45 (fixed) & $-1.76^{+0.13}_{-0.14}$ \\

\bottomrule

\end{tabular}
}
\end{table}

\begin{table}
\centering
    \caption{Best-fit double power-law parameters for the UV LF.}
    \label{tab:dpl_params}
\resizebox{\linewidth}{!}{
\begin{tabular}{cccccc}
\toprule

Redshift & $\logten(\Phi^*/{\rm Mpc}^{-3})$ & $M^*$ & $\beta$ & $\gamma$ \\

\midrule

2.76 - 2.99 & $-6.61^{+0.74}_{-0.45}$ & $-26.63^{+1.61}_{-0.75}$ & $-1.78^{+0.20}_{-0.33}$ & $-3.32^{+1.42}_{-0.91}$ \\
2.91 - 3.16 & $-6.75^{+0.89}_{-0.47}$ & $-26.74^{+2.09}_{-0.81}$ & $-1.80^{+0.20}_{-0.29}$ & $-3.04^{+1.72}_{-0.78}$ \\
3.08 - 3.32 & $-6.66^{+0.69}_{-0.49}$ & $-26.66^{+1.74}_{-0.94}$ & $-1.70^{+0.19}_{-0.28}$ & $-2.81^{+1.64}_{-0.60}$ \\
3.25 - 3.50 & $-6.83^{+0.68}_{-0.47}$ & $-26.64^{+1.76}_{-0.92}$ & $-1.65^{+0.22}_{-0.30}$ & $-2.91^{+1.69}_{-0.72}$ \\
3.41 - 3.66 & $-6.83^{+0.44}_{-0.44}$ & $-26.49^{+1.09}_{-0.87}$ & $-1.54^{+0.24}_{-0.30}$ & $-3.29^{+1.47}_{-0.96}$ \\
3.58 - 3.82 & $-6.92^{+0.58}_{-0.42}$ & $-26.69^{+1.52}_{-0.81}$ & $-1.60^{+0.23}_{-0.30}$ & $-3.15^{+1.62}_{-0.89}$ \\
\midrule
2.76 - 2.99 & $-5.92^{+0.21}_{-0.13}$ & $-25.23^{+0.65}_{-0.39}$ & -1.25 (fixed) & $-2.68^{+0.28}_{-0.27}$ \\
2.91 - 3.16 & $-5.92^{+0.43}_{-0.23}$ & $-24.89^{+1.30}_{-0.71}$ & -1.25 (fixed) & $-2.44^{+0.35}_{-0.25}$ \\
3.08 - 3.32 & $-6.00^{+0.38}_{-0.21}$ & $-25.06^{+1.24}_{-0.68}$ & -1.25 (fixed) & $-2.37^{+0.35}_{-0.24}$ \\
3.25 - 3.50 & $-6.31^{+0.32}_{-0.17}$ & $-25.47^{+1.07}_{-0.56}$ & -1.25 (fixed) & $-2.52^{+0.45}_{-0.34}$ \\
3.41 - 3.66 & $-6.58^{+0.17}_{-0.12}$ & $-26.00^{+0.56}_{-0.38}$ & -1.25 (fixed) & $-3.03^{+1.09}_{-0.56}$ \\
3.58 - 3.82 & $-6.54^{+0.25}_{-0.16}$ & $-25.89^{+0.85}_{-0.55}$ & -1.25 (fixed) & $-2.76^{+0.76}_{-0.45}$ \\

\bottomrule

\end{tabular}
}
\end{table}

\end{document}